\documentclass[lettersize,journal]{IEEEtran}
\usepackage{amsmath,amsfonts}
\usepackage{algorithmic}
\usepackage{algorithm}
\usepackage{array}
\usepackage[caption=false,font=normalsize,labelfont=sf,textfont=sf]{subfig}
\usepackage{textcomp}
\usepackage{stfloats}
\usepackage{url}
\usepackage{verbatim}
\usepackage{graphicx}
\usepackage{cite}
\usepackage[colorlinks=true, allcolors=blue]{hyperref}

\usepackage{cancel}
\usepackage{color}
\usepackage{ulem}    % for removed text (strikethrough)

\begin{document}

\title{Generating Reliable Initial Velocity Models for Full-waveform Inversion with Well and Structural Constraints}

\author{Qingchen Zhang, Shijun Cheng, Wei Chen and Weijian Mao
\thanks{Manuscript submitted Mar 3, 2025. This work was supported by National Natural Science Foundation of China (42274174). \it{(Corresponding author: Shijun~Cheng)}}
\thanks{Qingchen Zhang and Wei Chen are with Key Laboratory of Exploration Technology for Oil and Gas Resources of Ministry of Education, Yangtze University, Wuhan 430100, China; Cooperative Innovation Center of Unconventional Oil and Gas (Ministry of Education \& Hubei Province), Yangtze University, Wuhan 430100, China. (e-mail: qczhang@yangtzeu.edu.cn).}
\thanks{Shijun Cheng is with the Division of Physical Science and Engineering, King Abdullah University of Science and Technology, Thuwal 23955-6900, Saudi Arabia. (e-mail: sjcheng.academic@gmail.com).}
\thanks{Weijian Mao is with the Research Center for Computational and Exploration Geophysics, State Key Laboratory of Precision Geodesy, Innovation Academy for Precision Measurement Science and Technology, CAS,Wuhan 430077, China. (e-mail: wjmao@whigg.ac.cn).}
}

% The paper headers
\markboth{Submit to IEEE TRANSACTIONS ON GEOSCIENCE AND REMOTE SENSING, Mar~2025}%
{Shell \MakeLowercase{\textit{et al.}}: A Sample Article Using IEEEtran.cls for IEEE Journals}

\maketitle

\begin{abstract}
Full waveform inversion (FWI) plays an important role in velocity modeling due to its high-resolution advantages. However, its highly non-linear characteristic leads to numerous local minimums, which is known as the cycle-skipping problem. Therefore, effectively addressing the cycle-skipping issue is crucial to the success of FWI. Well-log data contain rich information about subsurface medium parameters, providing inherent advantages for velocity modeling. Traditional well-log data interpolation methods to build velocity models have limited accuracy and poor adaptability to complex geological structures. This study introduces a well interpolation algorithm based on a generative diffusion model (GDM) to generate initial models for FWI, addressing the cycle-skipping problem. Existing convolutional neural network (CNN)-based methods face difficulties in handling complex feature distributions and lack effective uncertainty quantification, limiting the reliability of their outputs. The proposed GDM-based approach overcomes these challenges by providing geologically consistent well interpolation while incorporating uncertainty assessment. Numerical experiments demonstrate that the method produces accurate and reliable initial models, enhancing FWI performance and mitigating cycle-skipping issues.

\end{abstract}

\section{Introduction}
\IEEEPARstart{F}{ull} waveform inversion (FWI) leverages the complete information contained in seismic waveforms, including amplitude, phase, and traveltime, to iteratively minimize the difference between recorded and simulated seismic data \cite{tarantola1984inversion,tarantola1986strategy}. Since having a theoretically highest resolution, FWI has been widely applied in fields such as geophysical exploration \cite{virieux2009overview}, medical imaging \cite{guasch2020full,xu2024full}, and engineering detection \cite{ernst2007full,meles2011gpr, tran2013sinkhole,rao2016guided}. Nevertheless, FWI is highly sensitive to the initial velocity model. An inaccurate initial model often causes the so-called cycle-skipping problem, where the inversion becomes trapped in local minima \cite{virieux2009overview,beller2018sensitivity,li2022well}. 

%As a result, developing robust strategies to obtain an appropriate initial velocity model remains a crucial challenge.

% The high-resolution FWI-provided subsurface parameter models enable researchers to better understand complex underground structures \cite{ernst2007full,feng2019multiscale}. In recent years, advancements in computational power and algorithms have significantly improved FWI techniques\cite{zhang2016robust,zhang2019elastic}. However, FWI has high computational complexity and strong dependency on the initial model (the cycle-skipping problem), making the rapid and accurate construction of the initial model a key focus of research \cite{virieux2009overview,beller2018sensitivity,li2022well}.

% The cycle-skipping problem leads to numerous local minima in the misfit function, particularly in multiparameter elastic scenarios. 

A common approach to mitigate cycle skipping involves adopting either multi-scale inversion algorithms \cite{bunks1995multiscale,boonyasiriwat2009efficient,ren2014multiscale,gao2019frequency} or global optimization methods. While multi-scale schemes attempt to incorporate low-frequency components first, they still fail when low-frequency data are absent. Global methods, exemplified by simulated annealing \cite{mendes2024faster}, can systematically explore the model space but often leads to high computational costs.

% Multiscale inversion algorithms \cite{bunks1995multiscale,boonyasiriwat2009efficient,ren2014multiscale,gao2019frequency} help mitigate this issue to some extent but still struggle when seismic data lack low-frequency information. Thus, constructing an accurate background model in the absence of low-frequency data becomes critical to resolving the cycle-skipping problem. Mendes et al. \cite{mendes2024faster} used simulated annealing to build initial velocity model for FWI, but the global algorithm would increase the computation cost.

Another promising strategy involves integrating FWI with alternative approaches to construct a good background velocity model, effectively forming a joint inversion. For example, Shin and Cha \cite{shin2008waveform}, Jun et al. \cite{jun2014laplace} demonstrated the reduced sensitivity of Laplacian-domain FWI to the low-frequency data and suggested that hybrid Laplacian–Fourier-domain FWI could solve the cycle-skipping problem to some degree. Sun et al. \cite{prieux2013building,sun2017alternating} employed ray-based traveltime tomography to build background velocity models for FWI. To overcome the high-frequency approximation limitations of ray-based tomography, Luo and Schuster \cite{luo1991wave} introduced correlation-norm wave-equation traveltime inversion. Ma and Hale \cite{ma2013wave} applied dynamic image warping (DIW) to estimate time shifts between recorded and synthetic data for wave-equation traveltime inversion. Luo et al. \cite{luo2016full} observed that earlier wave-equation-based traveltime inversion methods were inadvertently influenced by seismic amplitude information, so that they proposed a full traveltime inversion algorithm.

% On the other hand, many modified different-type objective functions are proposed to attempt to overcome the cycle-skipping problems of FWI. 
In addition to these joint inversion strategies, modifying the objective function itself can also helps to address cycle skipping issues. For example, Silva et al. \cite{da2022graph}, Engquist and Yang \cite{engquist2022optimal} proposed the optimal transport function to measure the discrepancy between observed and synthetic data by comparing their probability distributions. Compared to the conventional $L_2$ norm, the optimal transport objective function has better convexity, leading to a smoother optimization landscape with fewer local minima and reduced cycle skipping. \cite{hu2018retrieving} utilized the phase modulation/demodulation concept to provide the so-called  beat bone FWI. Besides, Chi et al. \cite{chi2014full}, Wu et al. \cite{wu2014seismic}, Wang et al. \cite{wang2018building}, and Hu et al. \cite{hu2022joint} developed envelope inversion methods to construct low-wavenumber velocity models without requiring low-frequency data. Warner and Guasch \cite{warner2016adaptive} proposed adaptive waveform inversion, introducing convolutional Wiener filters to iteratively match simulated and observed data. Reflection waveform inversion methods developed by Chi et al. \cite{chi2015correlation}, Guo and Alkhalifah \cite{guo2017elastic}, Wang et al. \cite{wang2020enhancing}, and Yao et al. \cite{yao2020review} enabled the separation of tomography and migration kernels, facilitating simultaneous reconstruction of low-wavenumber background velocities and high-wavenumber reflectivity using only reflected waves.

In recent years, various deep learning strategies have emerged as a powerful alternative to mitigate the cycle-skipping issue in FWI. The first category uses neural networks (NNs) to extrapolate missing low-frequency components, employing supervised \cite{ovcharenko2019deep, sun2020extrapolated, fang2020data}, semi-supervised \cite{sun2023learning}, or self-supervised techniques \cite{cheng2024self}. The second category applies purely data-driven methods to invert velocity models directly from observed data \cite{yang2019deep, li2019deep, zhang2021deep, du2022deep}. However, these approaches often have limited generalization when applied to geological settings different from those in the training dataset. The third category defines a NN-based objective function to measure the discrepancy between recorded and simulated data \cite{sun2020joint, saad2024siamesefwi}, but its ability to handle data with severe low-frequency gaps remains to be verified. The fourth category, known as implicit FWI \cite{sun2023implicit, zhang2023multilayer, sun2025implicit}, represents the velocity model through a network that starts from random initialization, thereby reducing dependence on an initial guess. Yet, this approach typically demands extensive iterations to produce a smooth inversion result and faces challenges when dealing with large-scale velocity models.

In this work, we propose a generative diffusion approach to mitigate cycle skipping in FWI by constructing a reliable initial velocity model. Our method learns the subsurface velocity prior from well data, migrated images, and initial models and, thus, uses that knowledge to construct more accurate starting points for inversion. To improve generalization, we outline a carefully designed dataset preparation strategy covering diverse geological scenarios. We also illustrate how to leverage the probabilistic nature of generative diffusion models (GDMs) to quantify the uncertainty in our constructed velocity models. Our approach is validated through both in-distribution and out-of-distribution model-building tests, and compared against a conventional well-log interpolation algorithm. The results indicate that our framework not only achieves superior and robust performance but also offers clearer uncertainty assessment in challenging geological conditions.

\section{Full-waveform inversion}
FWI is a data-fitting technique that aims to estimate subsurface parameters (e.g., seismic velocity) by minimizing the difference between observed and simulated seismic data \cite{virieux2009overview}. Under the acoustic approximation, the forward modeling of seismic data is often governed by the following wave equation:
\begin{equation}\label{eq_fwi_wave}
\nabla^2 p(\mathbf{x}, t; \mathbf{m})
- \frac{1}{v^2(\mathbf{x})} \frac{\partial^2 p(\mathbf{x}, t; \mathbf{m})}{\partial t^2}
= s(\mathbf{x}, t),
\end{equation}
where \(p(\mathbf{x}, t; \mathbf{m})\) is the acoustic pressure wavefield, \(\mathbf{x}\) denotes the spatial coordinates, \(t\) is time, \(v(\mathbf{x})\) is the seismic velocity (part of the model \(\mathbf{m}\)), and \(s(\mathbf{x}, t)\) represents the source term.

Let \(\mathbf{d}^{\text{obs}}\) be the observed seismic data and \(\mathbf{d}^{\text{syn}}(\mathbf{m})\) be the synthetic data generated by solving Equation~\eqref{eq_fwi_wave} under the model \(\mathbf{m}\). FWI seeks to minimize the misfit function
\begin{equation}\label{eq_fwi_obj}
\mathcal{J}(\mathbf{m})
= \frac{1}{2} \sum_{s}\sum_{r} \int \|\mathbf{d}^{\text{obs}}(r, s, t) - \mathbf{d}^{\text{syn}}(r, s, t; \mathbf{m})\|^2 \, dt,
\end{equation}
where \(s\) and \(r\) denote the source and receiver locations, respectively. This objective function measures the discrepancy between the observed and simulated data.

To find the model \(\mathbf{m}^{*}\) that minimizes \(\mathcal{J}\), one common approach is to use gradient-based optimization. Specifically, the gradient of \(\mathcal{J}\) with respect to the model parameters \(\mathbf{m}\) is computed via the adjoint-state method \cite{plessix2006review}, and the model is updated iteratively:
\begin{equation}\label{eq_fwi_update}
\mathbf{m}^{(k+1)}
= \mathbf{m}^{(k)}
- \alpha \,\nabla_{\mathbf{m}}\mathcal{J}\bigl(\mathbf{m}^{(k)}\bigr),
\end{equation}
where \(\alpha\) is a step size and \(k\) is the iteration index.

Although FWI can, in principle, recover high-resolution velocity models, its objective function is highly nonconvex. As a result, FWI can suffer from local minima, especially if the starting model is far from the true velocity distribution. This issue is often referred to as the “cycle-skipping” problem \cite{virieux2009overview}. 

Therefore, a key strategy for mitigating these challenges is to provide an good initial velocity model, thereby reducing the risk of cycle-skipping. Chen et al. \cite{chen2016geological} proposed a well log interpolation algorithm to construct an initial background model for FWI, utilizing the abundant low-wavenumber information in well log data. However, this conventional algorithm has a strong dependence on the completeness of the well log data. Meanwhile, when the subsurface structure is complex and there is no enough well logs, this algorithm might fail to build an accurate enough initial model for FWI. In the following, we address these limitations by integrating well data with structural constraints from migrated images within a generative diffusion framework. This integrated approach produces initial velocity models that capture both localized well information and large-scale geological structures, helping FWI avoid local minima and improving inversion accuracy.

\section{Method}
This section first provide an overview of GDMs, then describe how they are used to initial velocity model building (VMB), followed by details on our training dataset generation, and finally present the network architecture. 

% \subsection{FWI}
% FWI is a powerful technique in geophysics, particularly for seismic data analysis, where it is used to build high-resolution models of the subsurface velocity. The ability to utilize full-waveform data enables FWI reconstruct more accurate subsurface models, particularly in complex geological structures.

% However, the success of FWI depends heavily on the quality of the initial model and the ability to solve the inverse problem efficiently. One significant challenge in FWI is the cycle-skipping problem, where the inversion may converge to local minima, leading to inaccurate results, especially in the absence of low-frequency data. In the following paper, we introduce an GDM-based initial model building algorithm

\subsection{GDMs}\label{method_gdms}
The denoising diffusion probabilistic model (DDPM) \cite{ho2020denoising} is a generative model that learns to gradually add and then remove noise from data in order to generate realistic samples. The model consists of a forward process, where noise is added step by step, and a reverse process, where noise is removed to recover the original data.

In the forward process, noise is gradually added to the data sample $x_0$ at each step $t$ according to a Gaussian distribution:
\begin{equation}\label{eq1}
q(x_t | x_{t-1}) = \mathcal{N}(x_t; \sqrt{1 - \beta_t} \, x_{t-1}, \beta_t \mathbf{I})
\end{equation}
where $\beta_t$ is a variance parameter that controls the amount of noise added at each step, and $\mathbf{I}$ denotes the identity matrix, representing isotropic Gaussian noise. This noise addition process can be reparameterized to express $x_t$ directly in terms of the original sample $x_0$ and a noise term $\epsilon$:
\begin{equation}\label{eq2}
x_t = \sqrt{\bar{\alpha}_t} x_0 + \sqrt{1 - \bar{\alpha}_t} \epsilon
\end{equation}
where \( \epsilon \sim \mathcal{N}(0, I) \) represents Gaussian noise, and \( \bar{\alpha}_t = \prod_{i=1}^t (1 - \beta_i) \). Over \( T \) steps, this process transforms the data into a standard Gaussian distribution.

The reverse process aims to gradually remove the noise added during the forward process and recover the original data sample $x_0$. This process is modeled as a parameterized Markov chain that reverses the forward steps, starting from a Gaussian sample $x_T$ and iteratively predicting less noisy versions of the data at each timestep $t$.The reverse process is defined as:
\begin{equation}\label{eq3}
p_\theta(x_{t-1} | x_t) = \mathcal{N}(x_{t-1}; \mu_\theta(x_t, t), \Sigma_\theta(x_t, t)),
\end{equation}
where $\mu_\theta(x_t, t)$ and $\Sigma_\theta(x_t, t)$ are the predicted mean and variance of $x_{t-1}$ given $x_t$, parameterized by a neural network (NN) with learnable parameters $\theta$.

To reduce complexity and improve efficiency, the variance $\Sigma_\theta(x_t, t)$ is often fixed or predefined, allowing the NN to focus on learning the mean prediction $\mu_\theta(x_t, t)$. The mean is parameterized as:
\begin{equation}\label{eq4}
\mu_\theta(x_t, t) = \frac{1}{\sqrt{\alpha_t}} \left( x_t - \frac{1 - \alpha_t}{\sqrt{1 - \bar{\alpha}_t}} \epsilon_\theta(x_t, t) \right),
\end{equation}
where $ \epsilon_\theta(x_t, t)$ represents the NN’s estimate of the noise term $\epsilon$ added to the original data during the forward process, and $\alpha_t=1-\beta_t$. The variance is commonly defined as $\sigma_tz$, which is an optional noise term for stochastic sampling and $ z \sim \mathcal{N}(0, \mathbf{I})$. Therefore, the reverse process can be explicitly expressed as:
\begin{equation}\label{eq5}
x_{t-1} = \frac{1}{\sqrt{\alpha_t}} \left( x_t - \frac{1 - \alpha_t}{\sqrt{1 - \bar{\alpha}_t}} \epsilon_\theta(x_t, t) \right) + \sigma_t z.
\end{equation}

From Equation \ref{eq5}, we can see that the quality of the reverse process directly depends on the accuracy of the noise prediction. To enable the reverse process to accurately reconstruct the data, the network $\epsilon_\theta(x_t, t)$ must be trained to correctly predict the noise added at each time step $t$. This is achieved by minimizing the difference between the true noise $\epsilon$ and the model’s prediction $\epsilon_\theta(x_t, t)$. Instead of explicitly optimizing the likelihood of $p_\theta(x_0)$, the training objective simplifies to a noise prediction task:
\begin{equation}\label{eq6}
L = \mathbb{E}_{x_0, \epsilon, t} \left[ \| \epsilon - \epsilon_\theta(x_t, t) \|^2 \right],
\end{equation}
where $x_t$ is sampled using the forward process, and $t$ is randomly chosen during each training iteration. This loss function aligns the model’s prediction $\epsilon_\theta(x_t, t)$ with the true noise $\epsilon$, thereby ensuring that the reverse process can effectively remove noise step by step during sampling.

While predicting the noise $\epsilon$ is an effective approach, an alternative strategy is to directly predict the clean data $x_0$ at each step $t$. In this case, the network is trained to directly predict $x_0$ rather than the noise $\epsilon$. The corresponding loss function for $x_0$-prediction becomes:
\begin{equation}\label{eq7}
L = \mathbb{E}_{x_0, \epsilon, t} \left[ \| x_0 - x_{0, \theta}(x_t, t) \|^2 \right],
\end{equation}
where $x_{0, \theta}(x_t, t)$ is the model’s prediction of $x_0$ based on the noisy input $x_t$ and timestep $t$.

This $x_0$-prediction approach offers several advantages. First, it simplifies the reverse process by directly generating the target data rather than relying on intermediate noise predictions. Second, predicting $x_0$ reduces the accumulation of errors during iterative denoising, as the model explicitly learns to reconstruct the clean data. Third, empirical studies suggest that $x_0$-prediction leads to faster convergence and higher generation quality compared to noise prediction \cite{bansal2024cold}.

After we transform the prediction target to $x_0$, the reverse process can be represented as follows:
\begin{equation}\label{eq8}
x_{t-1} = \sqrt{\bar{\alpha}_{t-1}} x_{0,\theta}(x_t, t) + \sqrt{1 - \bar{\alpha}_{t-1}} \hat{\epsilon}(x_t, t)
\end{equation}
where $\hat{\epsilon}(x_t, t)$ is an estimate of the added noise as follows:
\begin{equation}\label{eq9}
\hat{\epsilon}(x_t, t) = \frac{x_t - \sqrt{\bar{\alpha}_t} x_{0,\theta}(x_t, t)}{\sqrt{1 - \bar{\alpha}_t}},
\end{equation}
which is deduced by Equation \ref{eq2}. Here, we can see that the predicted \( x_0 \) is used to refine \( x_{t-1} \). 

From Equation \ref{eq8}, it is evident that even if we sample an arbitrary \( x_t \), we can directly predict \( x_0 \) using the model \( x_{0,\theta}(x_t, t) \). This implies that we are not necessarily required to execute the reverse process step by step. Instead, we can directly map from a noisy input \( x_t \) to the clean data \( x_0 \), significantly speeding up the sampling process.

In our subsequent experiments, we further simplify the sampling process by starting from a random Gaussian noise $\epsilon \sim \mathcal{N}(0, \mathbf{I})$. Instead of performing $T$ iterative denoising steps, we directly use the last step of the reverse process to obtain the predicted $x_0$. This approach can be expressed as:
\begin{equation}\label{eq10}
x_0 = x_{0,\theta}(\epsilon, t=0),
\end{equation}
where $\epsilon$ is the initial noise sampled from a Gaussian distribution, and $x_{0,\theta}$ is the trained model that predicts the clean data directly from the noisy input.

This one-step sampling strategy greatly enhances efficiency while leveraging the predictive capability of $x_0$-targeting models. By skipping intermediate reverse steps, we can achieve faster sampling without compromising the quality of the generated results. This efficiency improvement is particularly critical in applications like VMB, where rapid generation of high-fidelity velocity models is essential.

\subsection{Initial seismic velocity model building with GDMs}\label{method_vmb}
Based on the powerful generation capabilities of the GDMs, we, here, use them to construct the initial velocity model for FWI. The core idea of this method is to utilize conditional constraints to control the generation of the initial velocity model, ensuring it aligns with the requirements of the inversion process, rather than merely synthesizing velocity models. These conditional constraints include well constraints, structural constraints, and an initial velocity model.

The structural constraints are derived from the migrated imaging results of the initial velocity model. While the imaging results are not fully accurate, they provide a general representation of the subsurface structures. By incorporating these constraints, we aim to generate velocity models that are better suited for FWI, helping to overcome local minima and improve inversion accuracy.

The conditional constraints are embedded directly into the input of the GDMs. Specifically, the input to the model consists of four channels, which can be represented as:
\begin{equation}\label{eq11}
x_\text{input} = (x_t, x_\text{init}, x_\text{well}, x_\text{migrated}).
\end{equation}
Here,
\begin{itemize}
    \item The first channel $x_t$ contains a noisy version of the true velocity model $x_0$, where we use Equation \ref{eq2} to corrupt true velocity model.
    \item The second channel $x_\text{init}$ includes the initial velocity model, serving as a baseline for the generated result.
    \item The third channel $x_\text{well}$ encodes well data, which is highly localized and sparse. This is represented as a 2D mask derived from the velocity model, where values outside the well locations are set to zero.
    \item The fourth channel $x_\text{migrated}$ contains the migrated imaging results of the initial velocity model, which offer structural information to guide the generation process.
\end{itemize}

The network is trained to directly predict the true velocity model $x_0$ from the noisy one $x_t$ and the corresponding conditional tensor $c=(x_\text{init}, x_\text{well}, x_\text{migrated})$. The training objective is formulated as:
\begin{equation}\label{eq12}
L = \mathbb{E}_{x_0, c, t} \left[ \|x_0 - x_{0,\theta}(x_t, c, t) \|^2 \right].
\end{equation}

After training, the model generates the initial velocity model through a reverse diffusion process. Starting from a random Gaussian noise $\mathbf{x}_T \sim \mathcal{N}(0, \mathbf{I})$, the velocity model $\mathbf{x}_0$ can be generated iteratively. However, since our network directly predicts $\mathbf{x}_0$, the sampling process can be simplified significantly in one step as:
\begin{equation}\label{eq13}
x_0 = x_{0,\theta}(\epsilon, c, t=0),
\end{equation}
where $c=(x_\text{init}, x_\text{well}, x_\text{migrated})$ represents the conditional input from the test data, and $x_{0,\theta}$ predicts the initial velocity model for the test data.

By embedding well, structural, and initial velocity constraints into the model, we ensure that the generated velocity models incorporate key prior knowledge, making them more suitable for seismic inversion tasks. The well constraints, though sparse, provide localized high-accuracy information, while the structural constraints from migrated imaging results guide the global subsurface structure. This approach helps to mitigate issues such as cycle skipping and local minima in FWI, ultimately enhancing the accuracy and reliability of the inversion results.

\subsection{Training data construction}\label{method_dataset}
To effectively train GDMs for generating initial velocity models, it is essential to minimize the gap between the synthetic velocity models used for training and the target velocity models in real-world applications. Synthetic velocity models could significantly differ from real-world counterparts, leading to potential generalization issues. To address this challenge, we need to carefully design synthetic velocity models by leveraging well data as key guidance, ensuring closer alignment with the target models.

The construction workflow, inspired by Ovcharenko et al. \cite{ovcharenko2022multi}, involves the following steps:

\begin{itemize}
   \item \textbf{Step 1: Generating random impedance sequences}: To emulate the distribution of impedance variations with depth, we start by creating a sparse sequence of random values uniformly sampled within the range of \(-1\) to \(1\). This forms the basis for generating subsurface heterogeneities.

   \item \textbf{Step 2: Constructing a dimensionless velocity profile}: The generated random sequence is integrated to form a dimensionless velocity profile, denoted as \( v(z) \). The values of \( v(z) \) are then normalized to fit within a predefined range from \( 0 \) to \( 1 \), ensuring consistency across the dataset.

   \item \textbf{Step 3: Extending to a 2D layered model}: The dimensionless velocity profile \( v(z) \) is extended laterally to create a 2D layered velocity model \( v(x, z) \), which is initially homogeneous along the horizontal direction. Well data is incorporated as localized constraints, ensuring that the generated model aligns with known subsurface properties.

   \item \textbf{Step 4: Introducing geological variability with elastic transforms}: To simulate realistic geological phenomena, the layered model \( v(x, z) \) is distorted using elastic transformations. These transformations introduce structural complexity, such as folding and intrusions, by applying 2D random Gaussian fields with adjustable mean and variance. The velocity values are then rescaled within a predefined range based on the expected target velocity properties.
\end{itemize}

By doing so, the final set of velocity models spans a wide range of structural variability, from relatively simple layered configurations to complex subsurface scenarios. These models are carefully designed to balance feature diversity and geological consistency, ensuring that the training dataset effectively captures the variability needed for robust model generalization.

Once the synthetic velocity models are constructed, we further generate migrated imaging results to provide structural constraints for GDM training. To ensure consistency with the target subsurface conditions, we align both the data acquisition system and the source wavelet used for generating seismic records. The alignment process involves the following steps:
\begin{itemize}
    \item Configuring source-receiver geometry, shot intervals, and offsets to replicate the acquisition system used for the target seismic data.
    \item Estimating a representative source wavelet for the target seismic data using traditional methods. For real data, techniques such as analyzing the amplitude spectrum, matching to synthetic seismograms, or leveraging well-log-derived reflectivity can provide a reasonable approximation of the source wavelet \cite{edgar2011reliable}.
    \item Using the aligned acquisition parameters and estimated source wavelet to generate synthetic seismic records based on the constructed velocity models.
    \item Performing migration on the synthetic seismic records to obtain imaging results consistent with those derived from the target velocity model.
\end{itemize}

Through the careful construction of synthetic velocity models and the alignment of acquisition systems and wavelets, the training dataset is designed to bridge the gap between synthetic and real data. Incorporating well data ensures localized accuracy, while simulating real-world acquisition systems and estimating realistic wavelets reduces systemic discrepancies. These strategies collectively aim to minimize generalization errors, allowing us to generate a good initial velocity models for FWI.

\subsection{Network architecture}\label{method_network}
\begin{figure}[!t]
\centering
\includegraphics[width=3.5in]{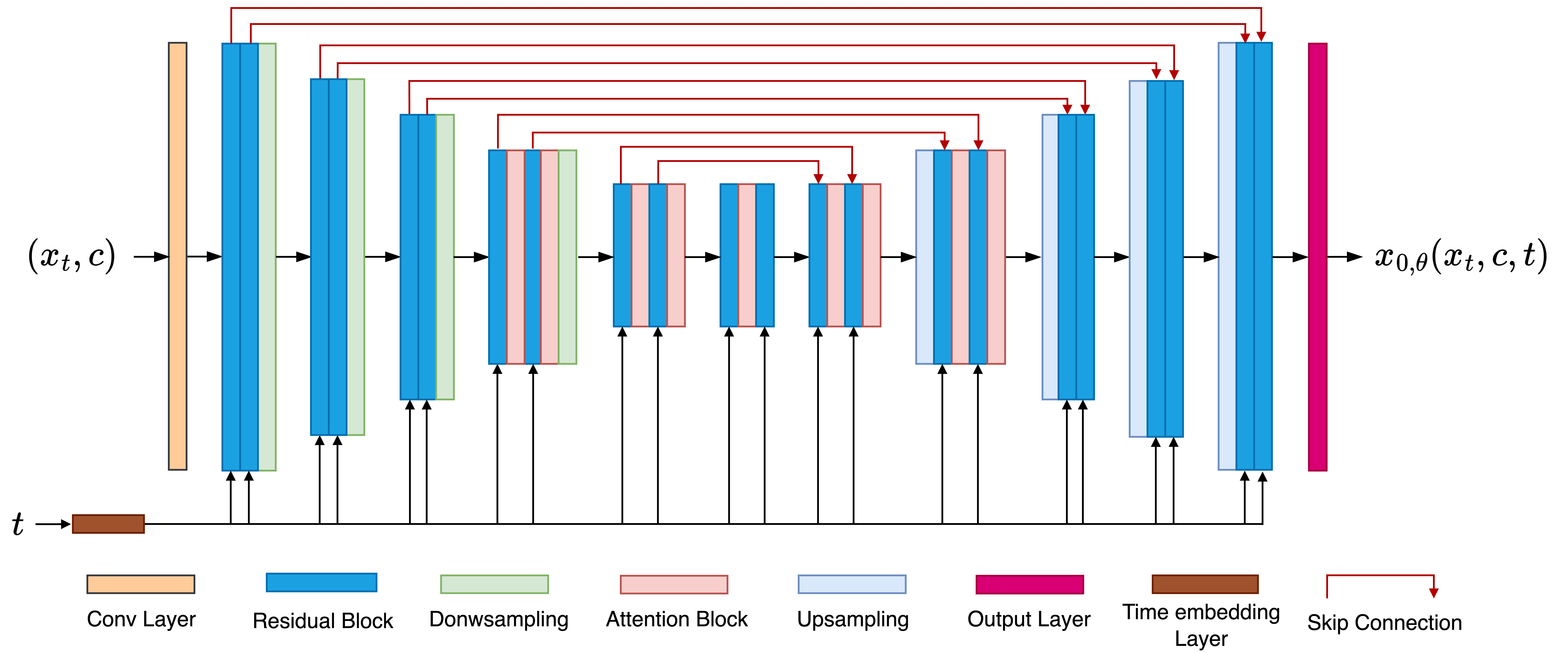}
\caption{An illustration of our network architecture. The input to the network is the noisy velocity model, $x_t$ and the conditions, $c$, and the output is the denoised velocity model, $x_0$, at time step, $t$. The components, including residual blocks, attention blocks, and time embedding layers, are described in Shi et al. \cite{diffobn2024}.}
\label{fig1}
\end{figure}

We adopt a popular U-Net-based architecture as shown in Fig. \ref{fig1}, specifically designed to generate initial velocity models for seismic inversion tasks. The architecture follows an encoder-decoder structure with skip connections to effectively preserve spatial features.

As previously described, the input $(x_t, c)$ to the network consists of a 4-channel tensor: the noisy velocity model, initial velocity model, well data, and migrated imaging result. The input tensor first passes through a 3$\times$3 convolutional layer, which adjusts the number of channels to 64, forming the base channel dimension for the network. The encoder progressively downsamples the feature maps to extract multi-scale representations. Each downsampling step consists of the following operations:
\begin{itemize}
    \item A residual block for learning effective feature representations.
    \item A downsampling operation achieved using a stride-2 convolution.
    \item The number of channels is doubled after each downsampling operation to increase the network's capacity to capture hierarchical features.
\end{itemize}

At the bottleneck layer, attention mechanisms are incorporated to capture long-range dependencies and improve the network's focus on important features. The attention block refines the encoded feature maps by highlighting critical regions in the data.

The decoder reconstructs the velocity model by gradually increasing the spatial resolution through upsampling. Each upsampling step includes:
\begin{itemize}
    \item A residual block to refine features.
    \item An upsampling operation implemented using transposed convolution.
    \item The number of channels is halved after each upsampling operation, symmetrically reversing the channel adjustments made in the encoder.
\end{itemize}
Skip connections are used to concatenate feature maps from corresponding encoder layers, ensuring that fine-grained spatial details are retained during reconstruction.
 
To condition the network on the diffusion time step $t$, a time embedding layer encodes the time step information into a learnable representation. This time embedding is injected into each residual block, enabling the network to adapt its predictions according to the diffusion step. Additionally, the conditional input $c$ (e.g., well data, initial velocity model, and migrated imaging result) is concatenated with the main input and integrated throughout the network.

The final output layer applies a 3$\times$3 convolution to map the reconstructed feature maps back to the target output, predicting the clean velocity model $x_0$. 

The network integrates key components, including residual blocks, attention blocks, and time embedding layers, as described in Shi et al. \cite{diffobn2024}. These modules collectively enhance the model's capability to capture hierarchical, spatial, and temporal features effectively.

\section{Numerical Examples}
In the following, we will present serval examples to evaluate the effectiveness of our proposed method for initial seismic velocity model building (VMB). First, we illustrate how the distribution of the constructed training dataset and training procedure. Next, we demonstrate our method on in-distribution velocity models to assess basic performance and consistency with the training data. We then validate the generalization capability of our approach on the well-known Marmousi model, which poses significant complexity. Finally, we further test our framework on the Overthrust model to examine its robustness under different geological scenarios. 

\subsection{Training data distribution and training procedure}
To create our training dataset, we start from a modified Marmousi model with a grid size of $300 \times 640$ and a grid spacing of $12.5\,\mathrm{m}$. As described in Section~\ref{method_dataset}, we extract three vertical profiles at horizontal locations of $1.25\,\mathrm{km}$, $3.75\,\mathrm{km}$, and $6.5\,\mathrm{km}$. These profiles serve as hypothetical well logs, which we use to construct multiple synthetic velocity models. Specifically, for each well log, we create 100 velocity realizations by applying the steps outlined in our velocity model construction approach.

\begin{figure}[htbp]
\centering
\includegraphics[width=3in]{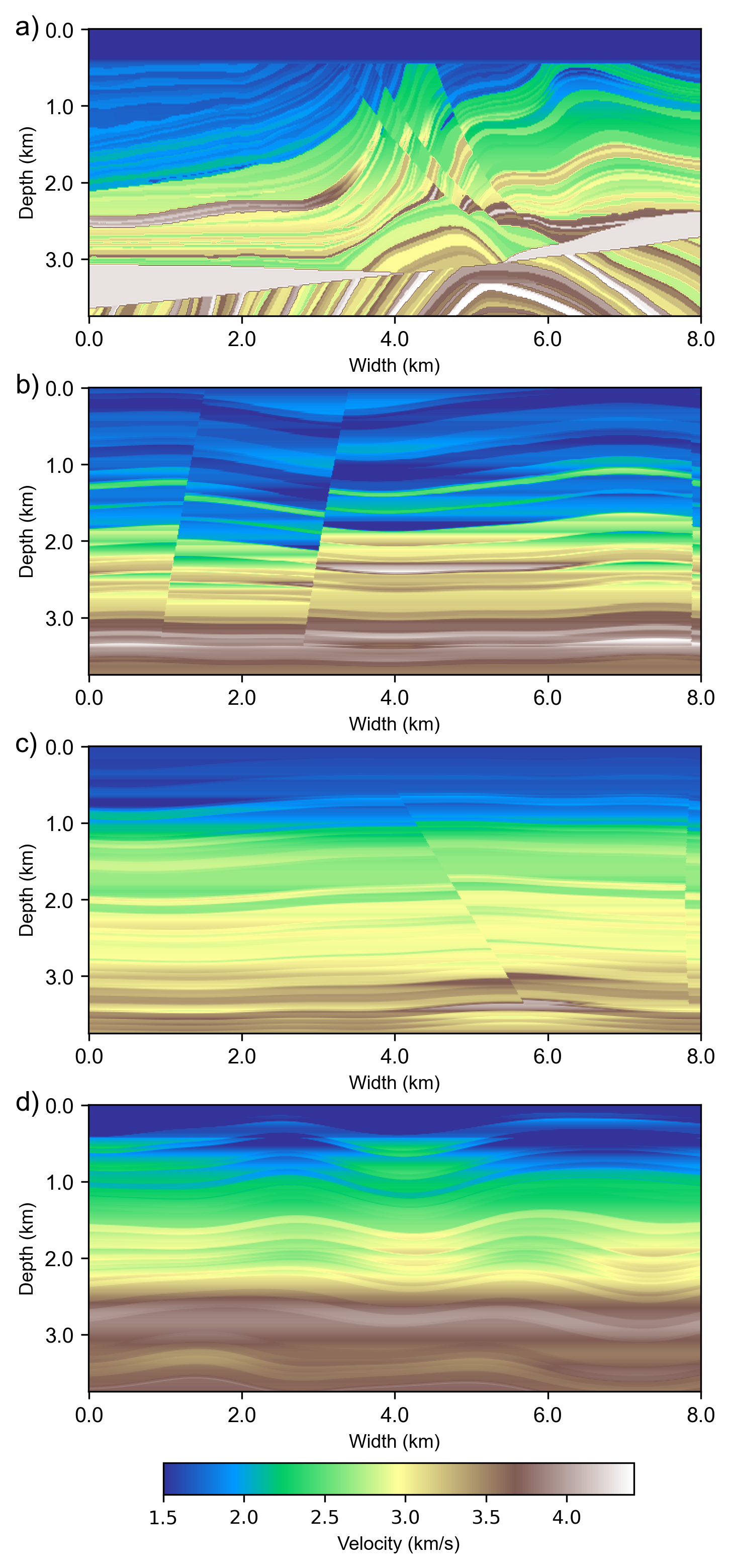}
\caption{(a) Modified Marmousi velocity model. (b), (c), and (d): Examples of synthetic velocity models constructed using the three extracted well profiles located at 1.25 km, 3.75 km, and 6.5 km, respectively.}
\label{fig2}
\end{figure}

Fig.~\ref{fig2}a shows the modified Marmousi model for reference, while Figs.~\ref{fig2}b-d illustrate representative examples of the synthetic velocity models generated from each of the three extracted well profiles. It should be noted that all the tests share the same acquisition system. We introduce a few artificial faults to simulate more complex subsurface structures, so that each set of synthetic models captures localized well constraints while exhibiting realistic lateral and vertical velocity variations.

\begin{figure}[htbp]
\centering
\includegraphics[width=3in]{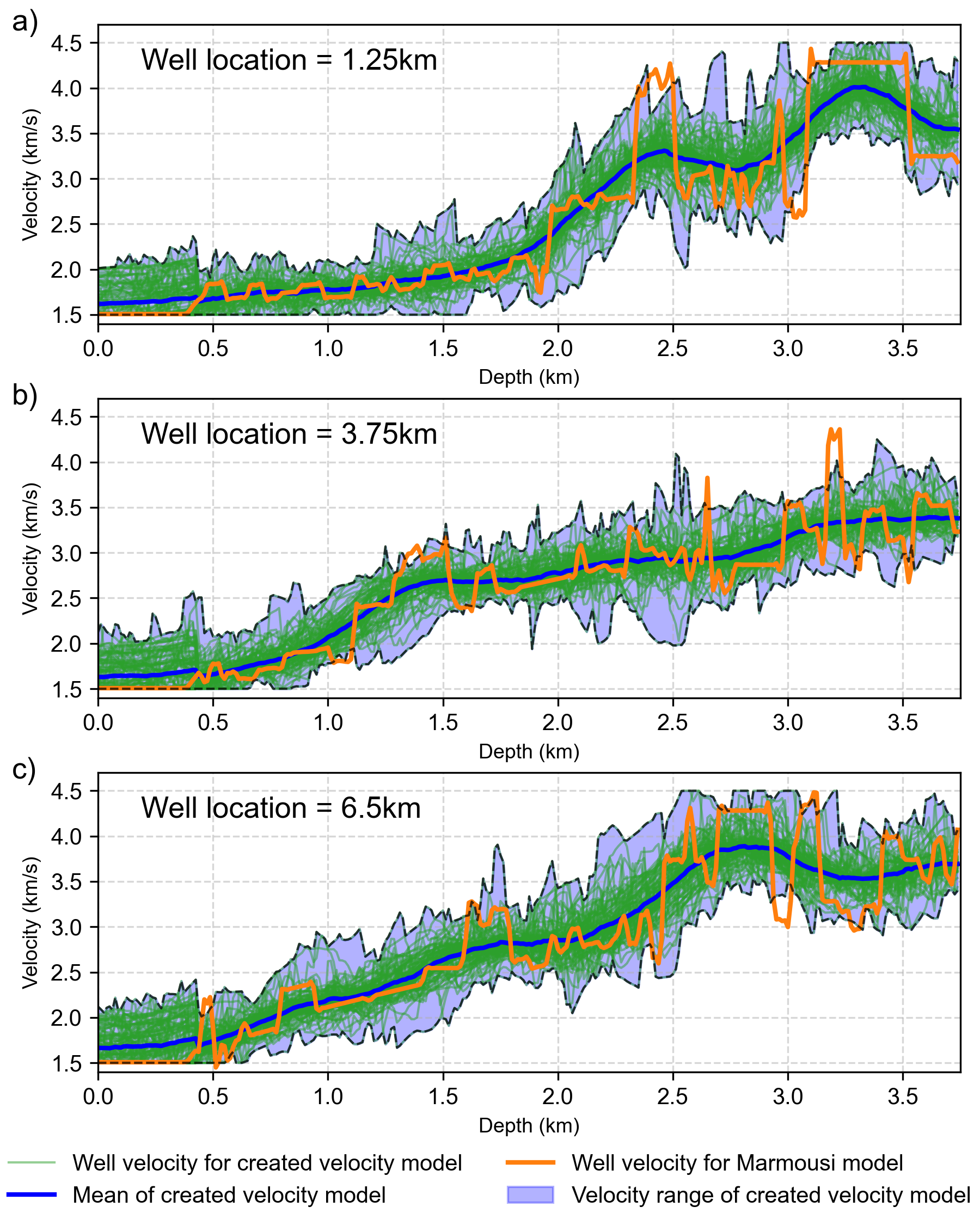}
\caption{Comparison of velocity profiles at each extracted well location. (a) Well at 1.25 km, (b) well at 3.75 km, and (c) well at 6.5 km. In each panel, the orange curve is the original Marmousi well-velocity log. The green curves show the 100 synthetic velocity profiles generated for that well location, 
and the blue curve is the mean of these 100 profiles. The shaded region indicates the range (minimum to maximum) across the 100 synthetic velocity models.}
\label{fig3}
\end{figure}

To assess the consistency between the synthetic models and the original Marmousi profiles, we plot their velocity curves in Fig.~\ref{fig3}. Panels~(a), (b), and (c) correspond to the three well locations at $1.25\,\mathrm{km}$, $3.75\,\mathrm{km}$, and $6.5\,\mathrm{km}$, respectively. In each panel, the orange curve denotes the extracted Marmousi log, while the green curves represent the 100 synthetic well profiles constructed for that location. We also compute the mean of these 100 profiles (blue curve) and shade the envelope between the minimum and maximum values. We can see that, the upper and lower bounds of the 100 synthetic velocity models at the corresponding well locations fully encompass the corresponding Marmousi well-velocity curve. Moreover, the mean of these 100 models closely follows the general trend of the Marmousi log, indicating that our constructed models capture the key velocity variations while allowing controlled perturbations. This diversity in the training set is critical for reducing overfitting in subsequent GDM training and for improving the robustness of the learned velocity distribution. 

For each created velocity model, we use acoustic propagator to simulate 106 shots, which are evenly distributed at the model surface and can be considered as our observed data. To consider a more strict observation condition, i.e. very limited offset, each shot contains only 120 geophones in single-side mode with an offset range from $0.25\ \mathrm{km}$ to $3.0\ \mathrm{km}$. Ricker wavelet with dominant frequency of $20\  \mathrm{Hz}$ is used as the source signal. We then use a larger degree of Gaussian smoothing ($\sigma=25$) on each of the constructed ground-truth velocity models to obtain our initial velocity models, and use acoustic RTM to produce the corresponding migration images. Thus, we obtain the 300 true velocity models, the initial velocity models, and the migration images derived from the initial velocity models required for the training dataset. Considering the scarcity of well data, we randomly select a single well from the true velocity model during each training iteration.

For our diffusion model training, we fix the learning rate at $1e-4$ and set the batch size to 5. We employ the AdamW optimizer \cite{loshchilov2017decoupled}, utilizing an exponential moving average with a decay rate of 0.999 to stabilize the training process. Training is conducted over 50000 iterations on a single NVIDIA A100 (80 GB) GPU, taking approximately 8 hours.

\subsection{Velocity model building (VMB) within the distribution}
To evaluate our method, we first present our test for velocity model construction within the distribution. Specifically, we use the well velocity at $x=1.25\,\mathrm{km}$ in the Marmousi model to build a synthetic test model, following the same procedure detailed in Section~\ref{method_dataset}. This synthetic example is chosen so that its overall statistical properties drawn from a similar distribution as our training data. 

\begin{figure}[htbp]
\centering
\includegraphics[width=3in]{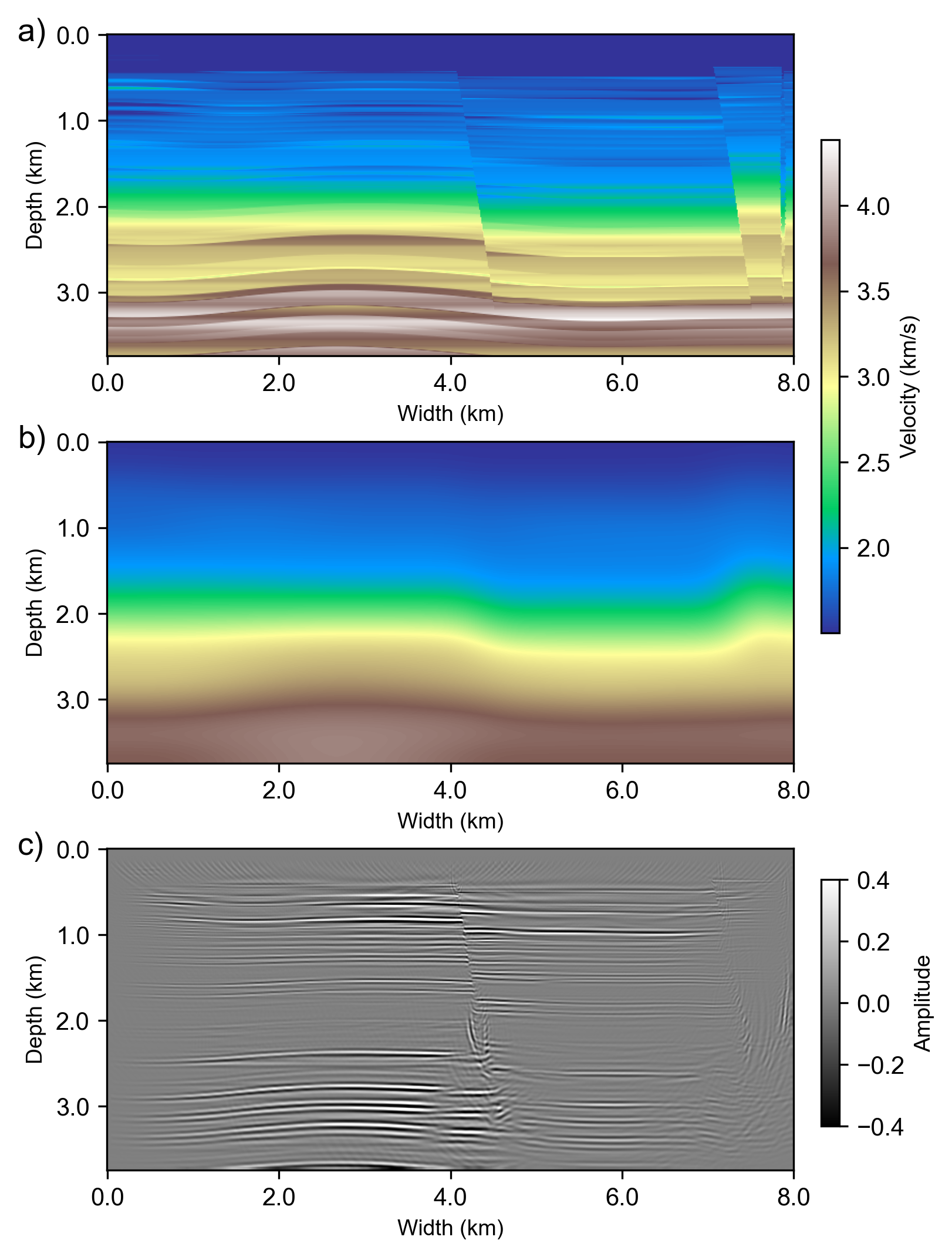}
\caption{(a) Synthetic velocity model constructed from a single Marmousi well at $x=1.25\,\mathrm{km}$. (b) Smoothed version of the synthetic model. (c) RTM image obtained with the initial (smoothed) velocity. }
\label{fig4}
\end{figure}

Fig.~\ref{fig4}a shows this synthetic velocity model, while Fig.~\ref{fig4}b presents the smoothed initial velocity model derived from it. We then perform reverse-time migration (RTM) using this initial model and display the resulting subsurface image in Fig.~\ref{fig4}c. 
Despite being generated with the initial velocity, the RTM result is able to capture the general structural features, such as layers and faults. This structural information, although imperfect, can serve as a valuable prior for our VMB framework.

\begin{figure}[htbp]
\centering
\includegraphics[width=3in]{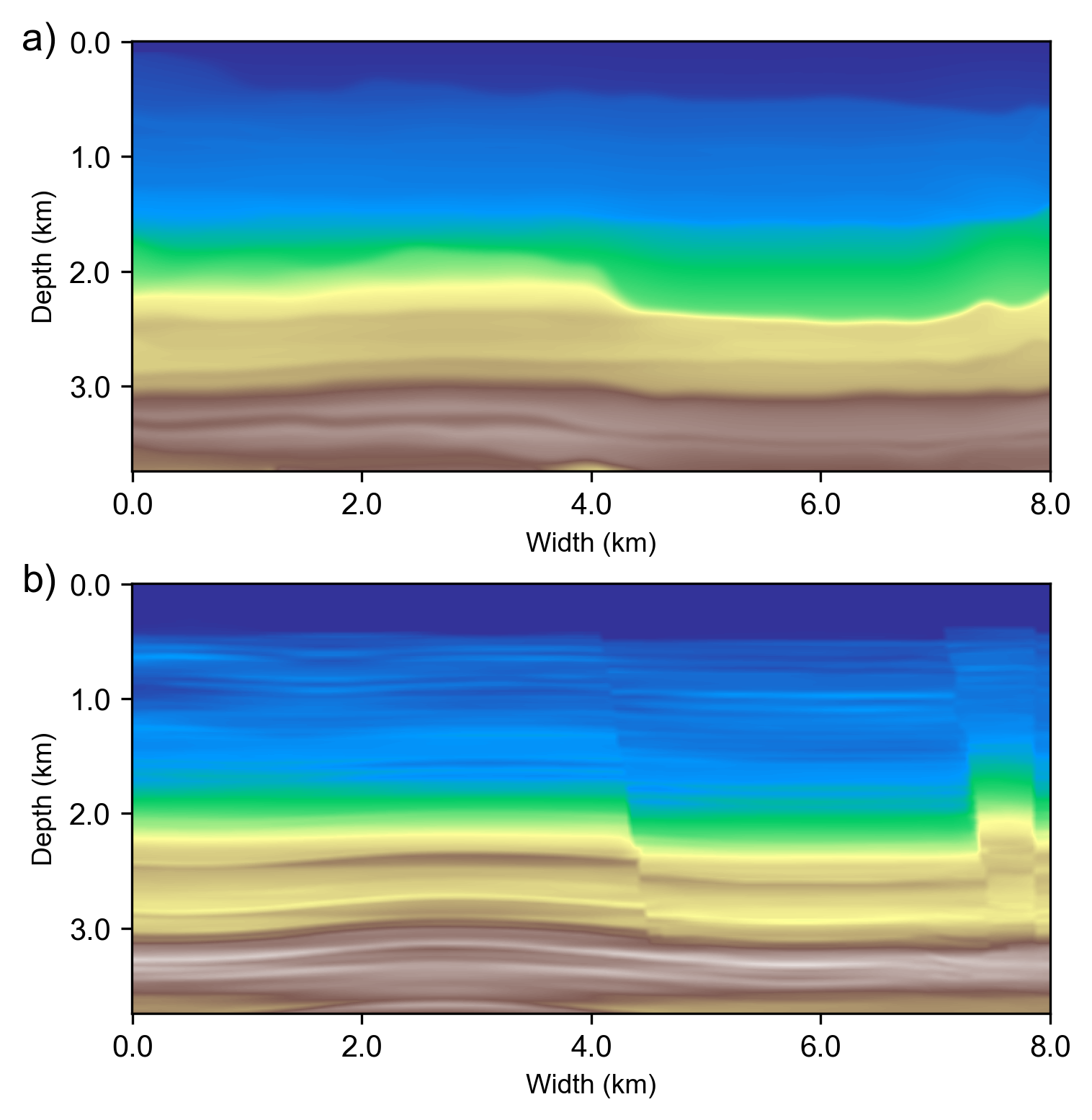}
\caption{Initial velocity models constructed for synthetic velocity model using two different approaches: (a) CVMBM \cite{chen2016geological,li2022well}, (b) our method. The color scale matches that of Fig.~\ref{fig4}.}
\label{fig5}
\end{figure}

In practical scenarios, only a limited number of wells are usually available. Therefore, our approach is restricted to acquiring a single well in both this and the subsequent tests. The well is positioned at a horizontal distance of $x = 1.25$ km in the true velocity model. To make a comparison, we use the conventional VMB method \cite{chen2016geological,li2022well} (well interpolation with plane wave destruction algorithm, hereafter CVMBM) as our benchmark. Figs.~\ref{fig5}a and \ref{fig5}b show the initial models generated by CVMBM and our method, respectively (the color scale matches that in Figs.~\ref{fig4}a and \ref{fig4}b). It is evident that CVMBM yields a more smoothed model, offering an approximate representation of the velocity distribution but failing to capture certain complex structures, including near-vertical faults. In contrast, our method produces a more detailed velocity field, accurately outlining the fault zones and providing higher resolution overall.

\begin{figure}[htbp]
\centering
\includegraphics[width=3in]{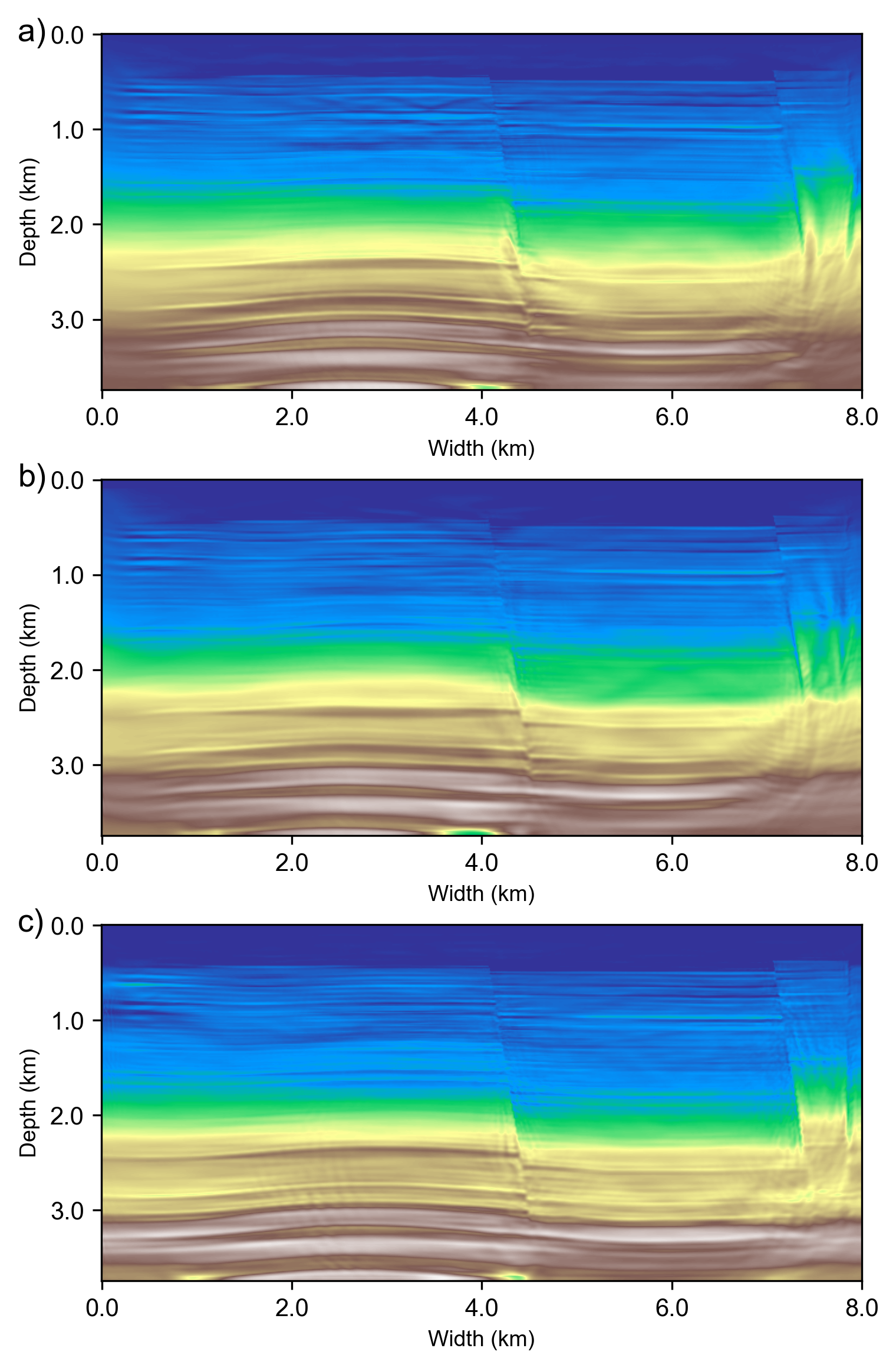}
\caption{FWI results for the synthetic model using three different initial velocity models. From top to bottom: (a) the original smoothed model, (b) the CVMBM-created model, (c) the model from our approach. The color scale matches that of Fig.~\ref{fig4}.}
\label{fig6}
\end{figure}

We then carry out full-waveform inversion (FWI) using three different starting models: (1) the original smoothed initial model, (2) the interpolated initial model given by CVMBM, and (3) the generated initial model produced by our method. Because real seismic data often lack frequencies below $5\,\mathrm{Hz}$, we here, as well as the following test, apply a low-cut filter at $5\,\mathrm{Hz}$ to simulate realistic field conditions. 
Fig.~\ref{fig6} compares the final FWI results (top to bottom) obtained from these three initial models. Neither the original smooth model nor the CVMBM-based model successfully recovers the vertical fault zone, due to their inability to provide a suitable structural prior. By contrast, our approach preserves higher-resolution features in the starting model, enabling FWI to resolve the fault zone more accurately.

\begin{figure}[htbp]
\centering
\includegraphics[width=3in]{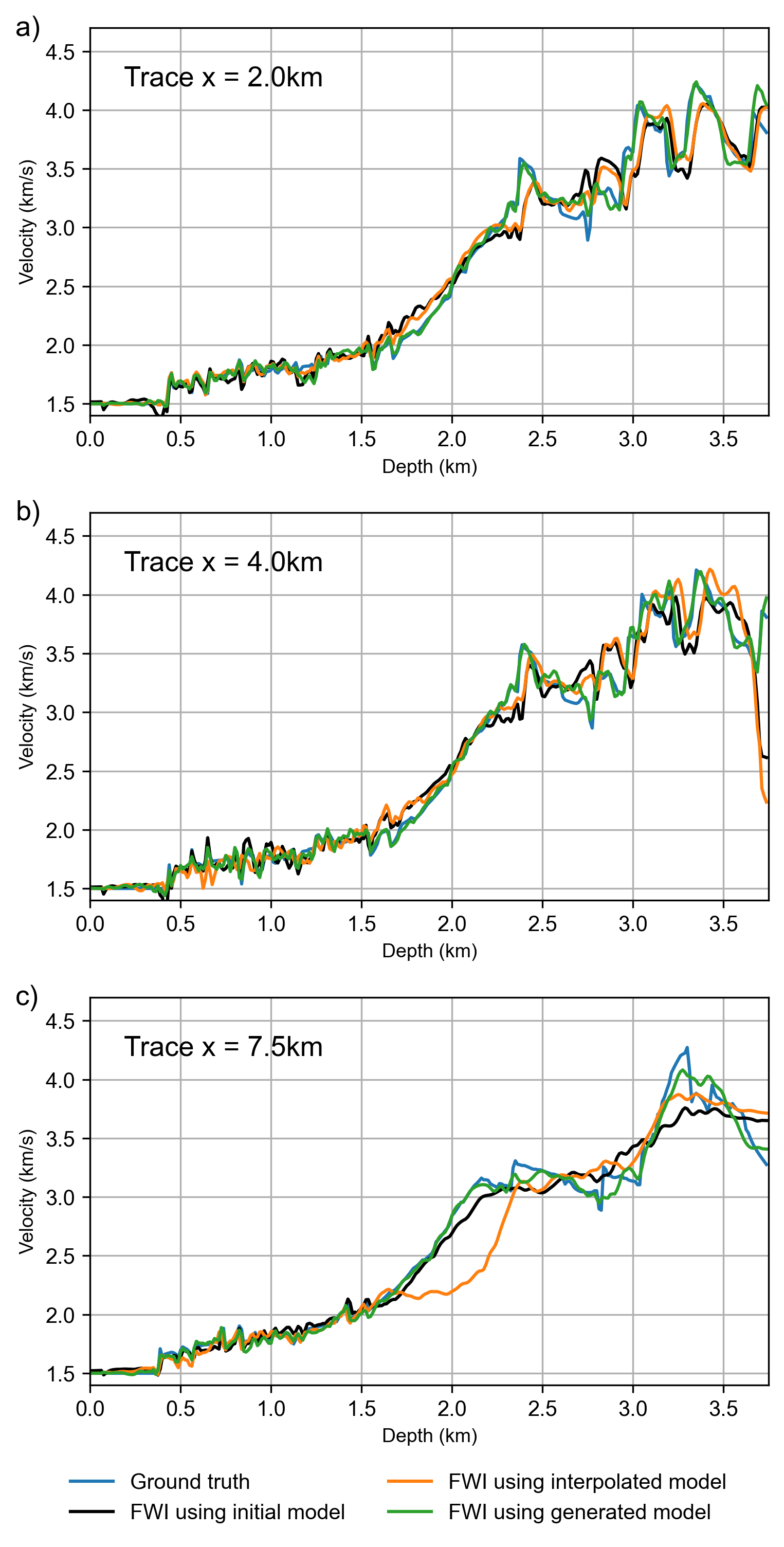}
\caption{Comparison of vertical velocity profiles from the synthetic velocity model and the three FWI results at three horizontal distances: (a) $x=2.0\,\mathrm{km}$, (b) $x=4.0\,\mathrm{km}$, (c) $x=7.5\,\mathrm{km}$.}
\label{fig7}
\end{figure}

To further quantify the inversion accuracy, we extract three vertical velocity profiles at $x=2.0\,\mathrm{km}$, $4.0\,\mathrm{km}$, and $7.5\,\mathrm{km}$ from the ground-truth model and each final FWI result (see Fig.~\ref{fig7}a-c). Our method shows the best agreement with the true velocity, especially around the faulted regions (e.g., at $4.0\,\mathrm{km}$ and $7.5\,\mathrm{km}$), while the initial smooth model and the CVMBM-created model lead to larger velocity errors at those depths.

\subsection{Marmousi model}
\begin{figure}[htbp]
\centering
\includegraphics[width=3in]{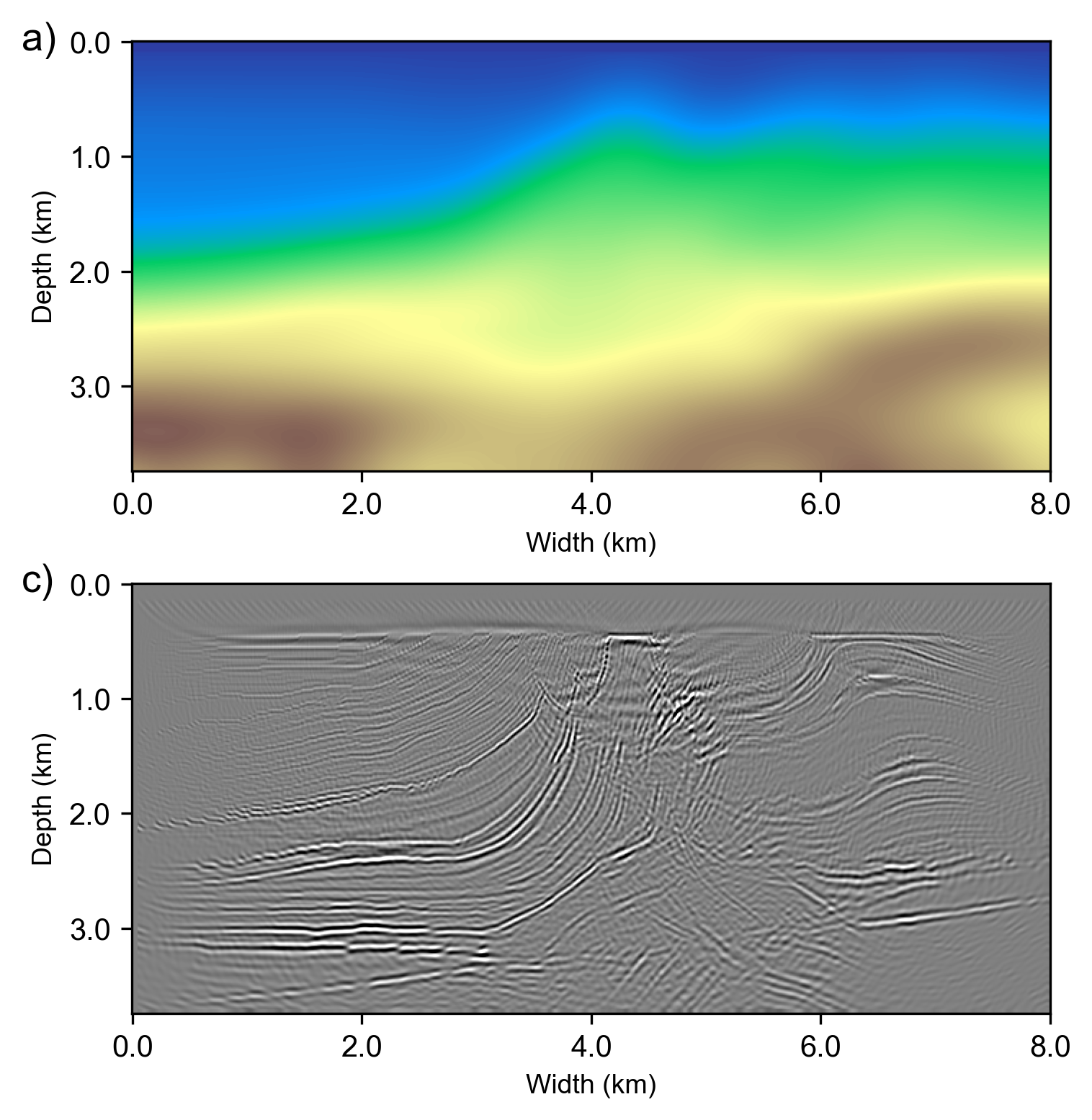}
\caption{(a) Initial velocity model obtained by smoothing the Marmousi model (shown in Fig. \ref{fig2}a). (b) RTM image computed using the smoothed velocity model in (a).}
\label{fig8}
\end{figure}

In our second numerical test, we assess the performance of our method on the classic Marmousi model. Fig.~\ref{fig8}a shows the initial velocity model obtained by smoothing the true Marmousi velocity field. Following the same strategy as before, we then perform RTM using this smoothed model. The corresponding subsurface image is shown in Fig.~\ref{fig8}(c). Although this migration uses a simplified starting model, it still highlights major layer boundaries. However, we must admit that it fails to effectively image some complex fault structures.

\begin{figure}[htbp]
\centering
\includegraphics[width=3in]{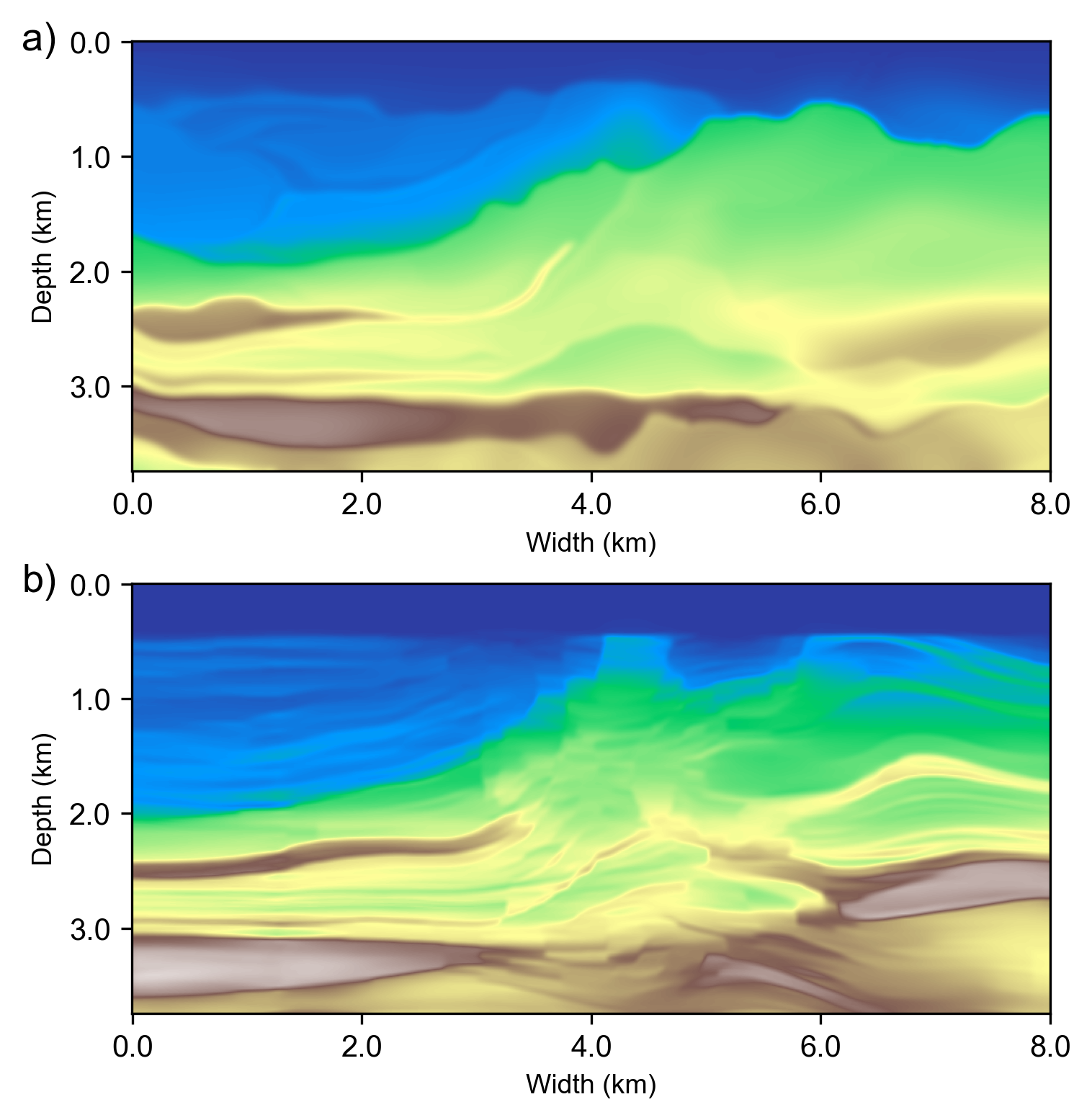}
\caption{Initial velocity models constructed for Marmousi model using two different approaches: (a) CVMBM, (b) our method.}
\label{fig9}
\end{figure}

Next, we construct two refined initial models: one using the CVMBM approach (Fig.~\ref{fig9}a) and the other with our method (Fig.~\ref{fig9}b). We again observe that CVMBM yields an overall smooth velocity distribution but tends to blur complex features, 
whereas our model preserves sharper details.

\begin{figure}[htbp]
\centering
\includegraphics[width=3in]{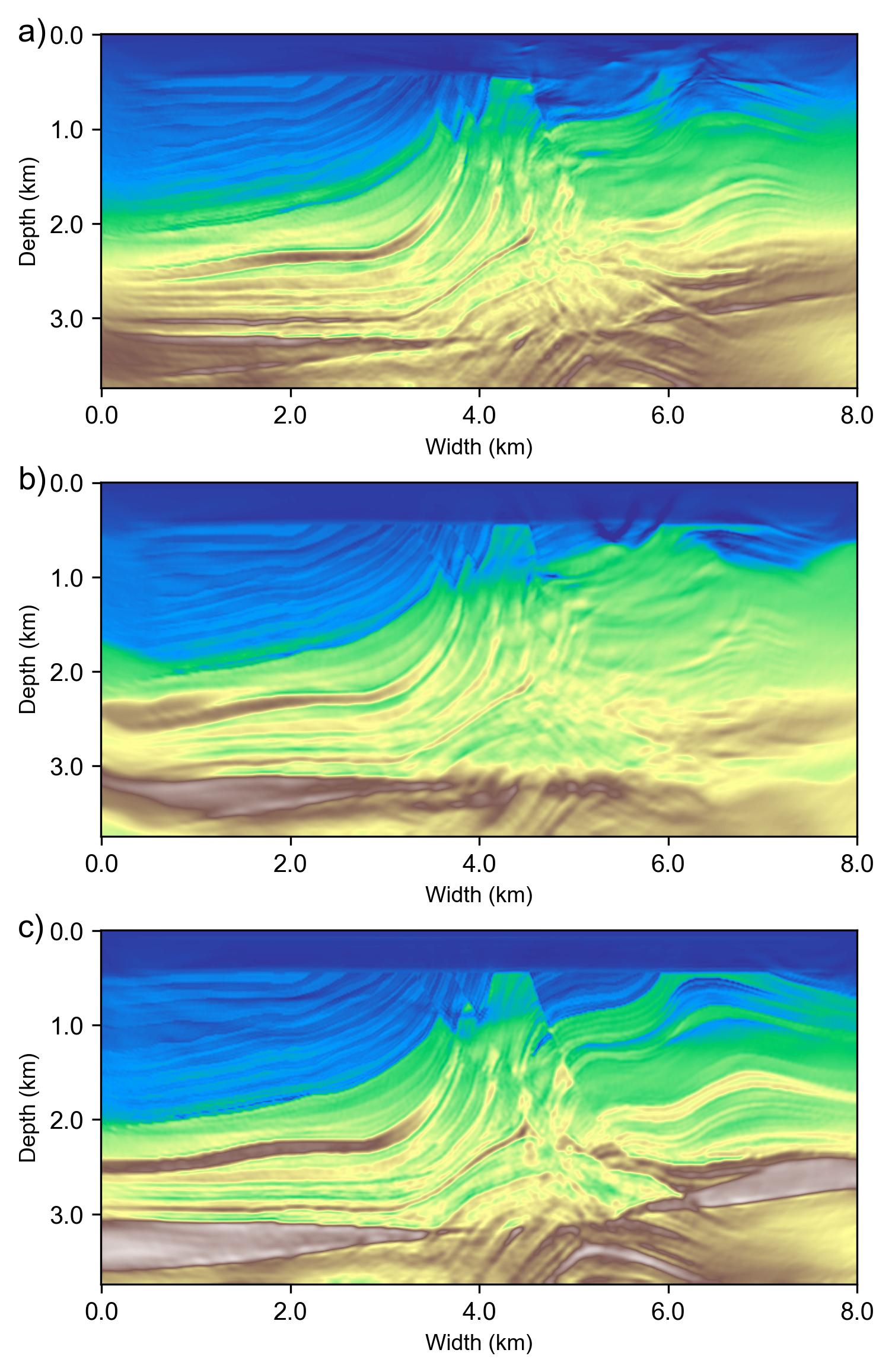}
\caption{FWI results for Marmousi model using three different initial velocity models. From top to bottom: (a) the original smoothed model, (b) the CVMBM-created model, (c) the model from our approach.}
\label{fig10}
\end{figure}

\begin{figure}[htbp]
\centering
\includegraphics[width=3in]{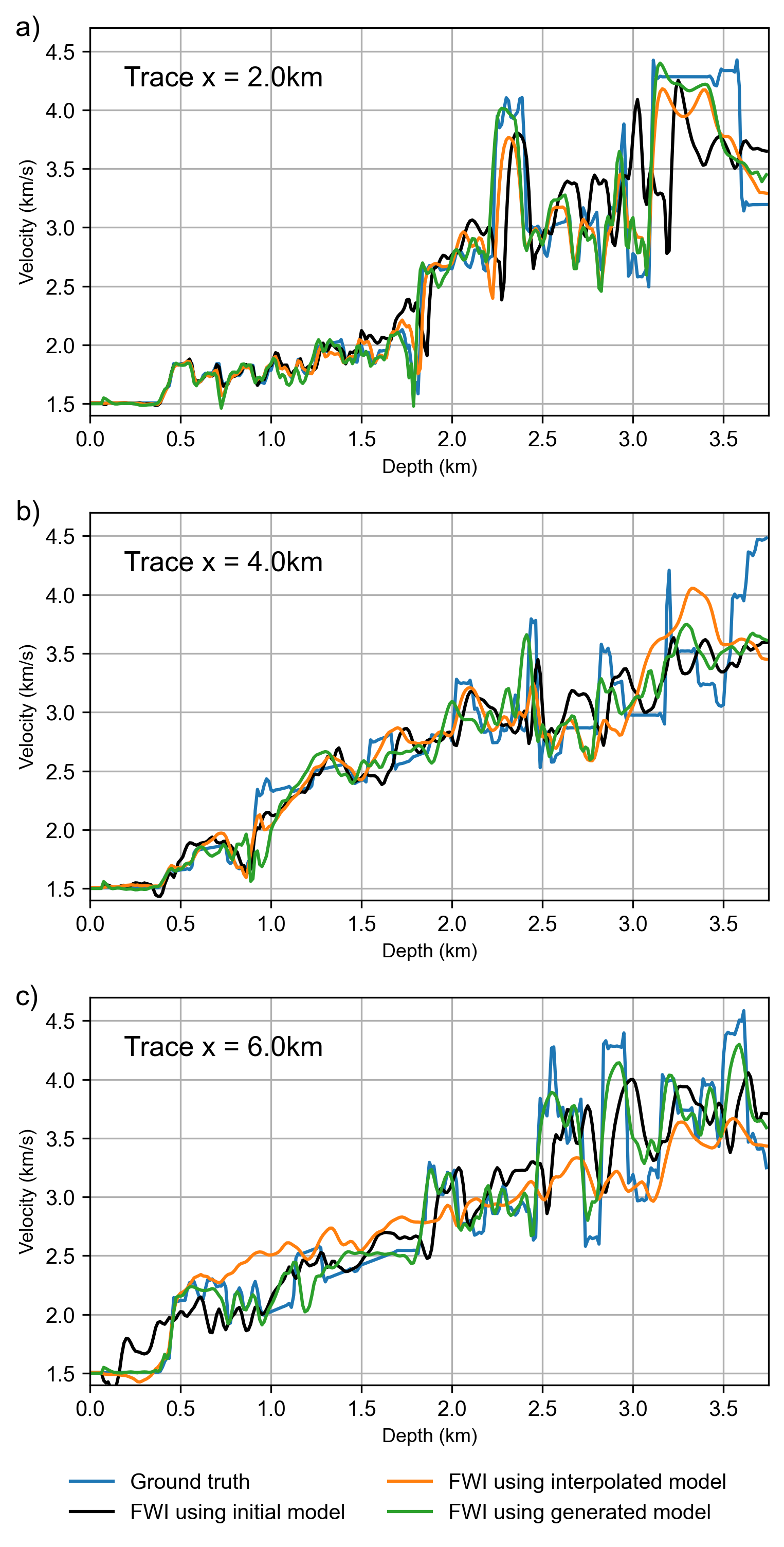}
\caption{Comparison of vertical velocity profiles from the true Marmousi model and the three FWI results at three horizontal distances: (a) $x=2\,\mathrm{km}$, (b) $x=4\,\mathrm{km}$, (c) $x=6\,\mathrm{km}$. }
\label{fig11}
\end{figure}

To evaluate the impact of these initial models, we run FWI with each starting point and compare the final velocity models in Fig.~\ref{fig10} (top to bottom). As in previous examples, the more accurate model information from our approach leads to superior reconstruction of subsurface complexities. Fig.~\ref{fig11} further demonstrates this advantage by comparing vertical velocity profiles at $x=2\,\mathrm{km}$, $4\,\mathrm{km}$, and $6\,\mathrm{km}$. 
Our FWI result (green curve) closely matches the true Marmousi velocity (blue curve), especially around faulted regions, while the smoother initial model (black curve) and the CVMBM-based model (orange curve) exhibit notable discrepancies.

\subsection{Overthrust model}
\begin{figure}[htbp]
\centering
\includegraphics[width=3in]{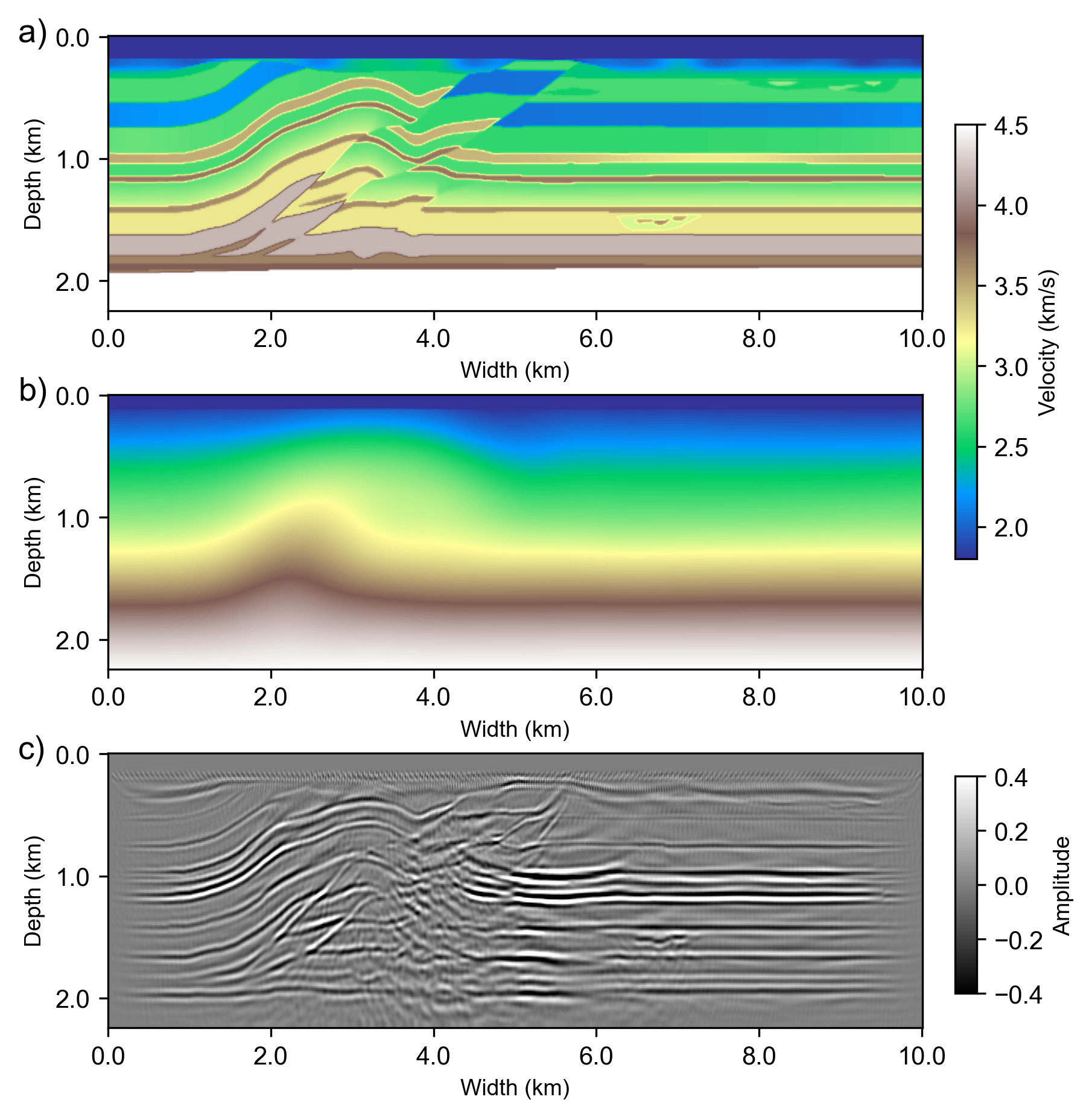}
\caption{(a) Overthrust velocity model. (b) Smoothed version of the true velocity model. (c) RTM image obtained with the initial (smoothed) velocity.}
\label{fig12}
\end{figure}

Our final experiment assesses the generalization of the proposed framework on the Overthrust model, which lies outside the training data distribution. The model has a grid size of $180 \times 801$ and a grid spacing of $12.5\,\mathrm{m}$. Fig.~\ref{fig12}a shows the true Overthrust velocity model, while Fig.~\ref{fig12}b presents the smoothed initial velocity derived from it. A total of 160 shots, evenly distributed at the surface, are excited in the following tests. Also, each shot is recorded by 120 geophones with an offset range from $0.25\ \mathrm{km}$ to $3.0\ \mathrm{km}$, and the source sginal is a Ricker wavelet with dominant frequency of $15.0\ \mathrm{Hz}$.  We then perform RTM with this initial model and display the resulting image in Fig.~\ref{fig12}c. The migrated image reveals major layered structures but fails to position the fault accurately due to velocity inaccuracies. This poses a challenge for the subsequent VMB process.

\begin{figure}[htbp]
\centering
\includegraphics[width=3in]{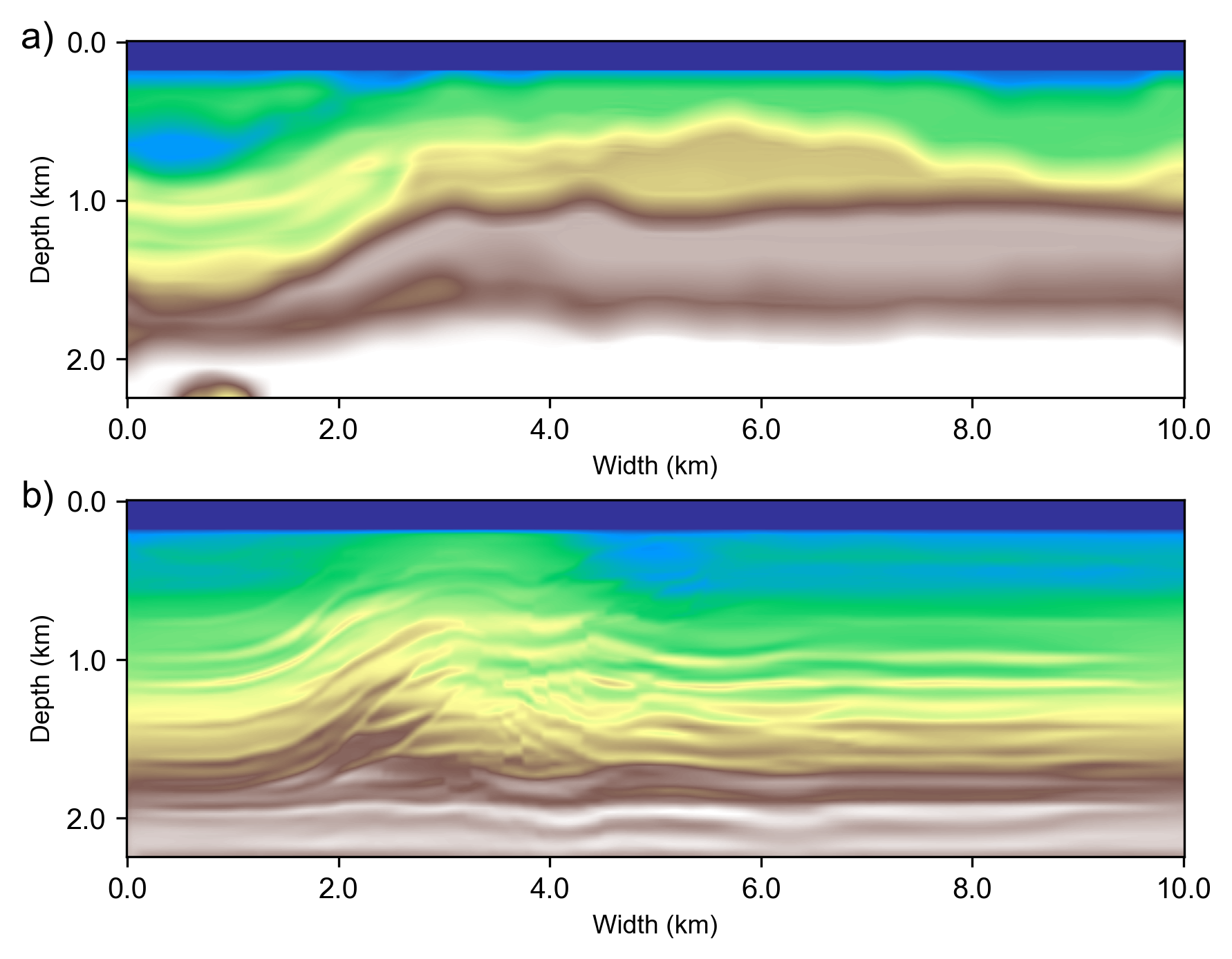}
\caption{Initial velocity models constructed for Overthrust model using two different approaches: (a) CVMBM \cite{chen2016geological,li2022well}, (b) our method.}
\label{fig13}
\end{figure}

Given the migrated image, as well as the single well (at $x=1.25\,\mathrm{km}$), we construct two refined starting models for WI using CVMBM (Fig.~\ref{fig13}a) and our approach (Fig.~\ref{fig13}b). While CVMBM provides a reasonable velocity distribution on the left side of the model, it neither accurately captures the central overthrust fault nor preserves smaller-scale features, 
ultimately yielding an overly smoothed representation. In contrast, our method offers a higher-quality initial model, more precisely building the faults in the center area, which is a critical factor for successful FWI.

\begin{figure}[htbp]
\centering
\includegraphics[width=3in]{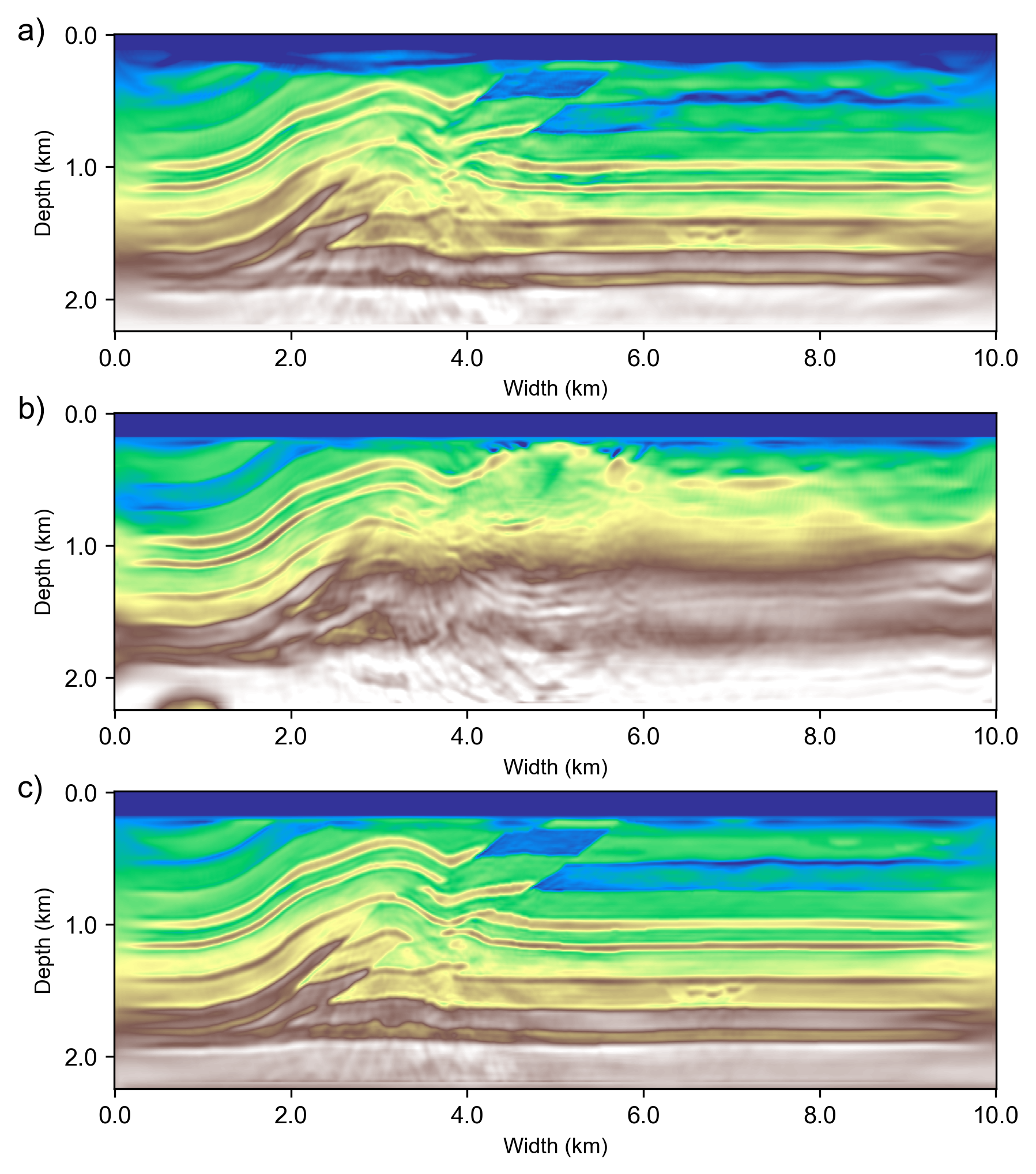}
\caption{FWI results for Overthrust model using three different initial velocity models. From top to bottom: (a) the original smoothed model, (b) the CVMBM-created model, (c) the model from our approach.}
\label{fig14}
\end{figure}

Fig.~\ref{fig14} compares final FWI results for three different initial models: the original smoothed velocity, the CVMBM-based model, and our generated model (top to bottom). Notably, the original smoothed model leads to a more accurate inversion than CVMBM. This indicates that CVMBM can struggle under conditions of sparse well data, leading to a less robustness to the initial model and potentially inducing a poorer FWI outcome. Our approach, on the other hand, yields a significant accuracy boost, recovering finer-scale layers and small structural elements that enhance the overall resolution of the inversion.

\begin{figure}[htbp]
\centering
\includegraphics[width=3in]{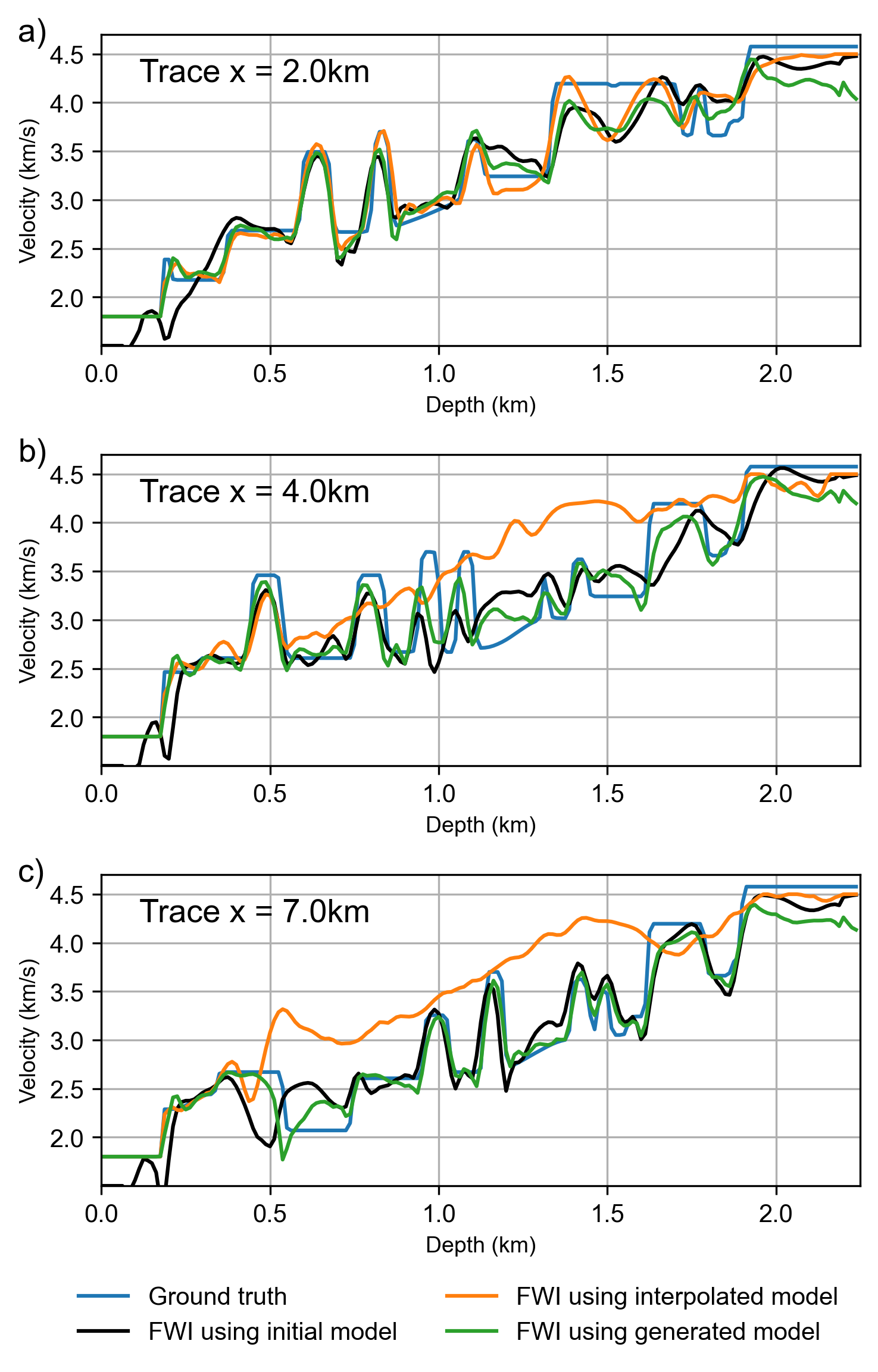}
\caption{Comparison of vertical velocity profiles from the true model and three FWI results at three horizontal distances: (a) $x=2\,\mathrm{km}$, (b) $x=4\,\mathrm{km}$, (c) $x=7\,\mathrm{km}$.}
\label{fig15}
\end{figure}

To further illustrate the agreement between the inverted and true velocities, we extract three profiles at $x=2\,\mathrm{km}$, $4\,\mathrm{km}$, and $7\,\mathrm{km}$ (Fig.~\ref{fig15}a-c). In all three panels, our method (green curve) aligns best with the ground truth, whereas the smoothed model (black) shows moderate accuracy, and the CVMBM-induced inversion (orange) diverges significantly in the middle and right portions of the domain. These results confirm that incorporating well data and structural constraints into our framework 
yields robust initial velocity models that enable more reliable FWI, even under challenging out-of-distribution conditions.

\section{Discussion}
In the following discussion, we first highlight how our approach performs uncertainty quantification during velocity model building (VMB) process, then address its current limitations, and finally propose directions for future research. 
\subsection{Velocity model building with uncertainty quantification}
One of the key advantages of using a generative diffusion model (GDM) for VMB is its natural ability to quantify uncertainty. Some studies have leveraged GDMs to assess the uncertainty in seismic processing outputs, thereby supporting decision-making \cite{diffobn2024, cheng2025agenerative}. Similarly, during the sampling phase for generating the velocity model, we can replicate the conditional inputs (initial model, well, migration results) $B$ times along the batch dimension and then draw $B$ distinct random noise samples. This yields $B$ realizations of the velocity model, each of which holds a minor difference. By averaging the $B$ generated velocity models, we obtain our final velocity model. Meanwhile, calculating the standard deviation allows us to assess the uncertainty of the constructed velocity model.

\begin{figure}[htbp]
\centering
\includegraphics[width=3in]{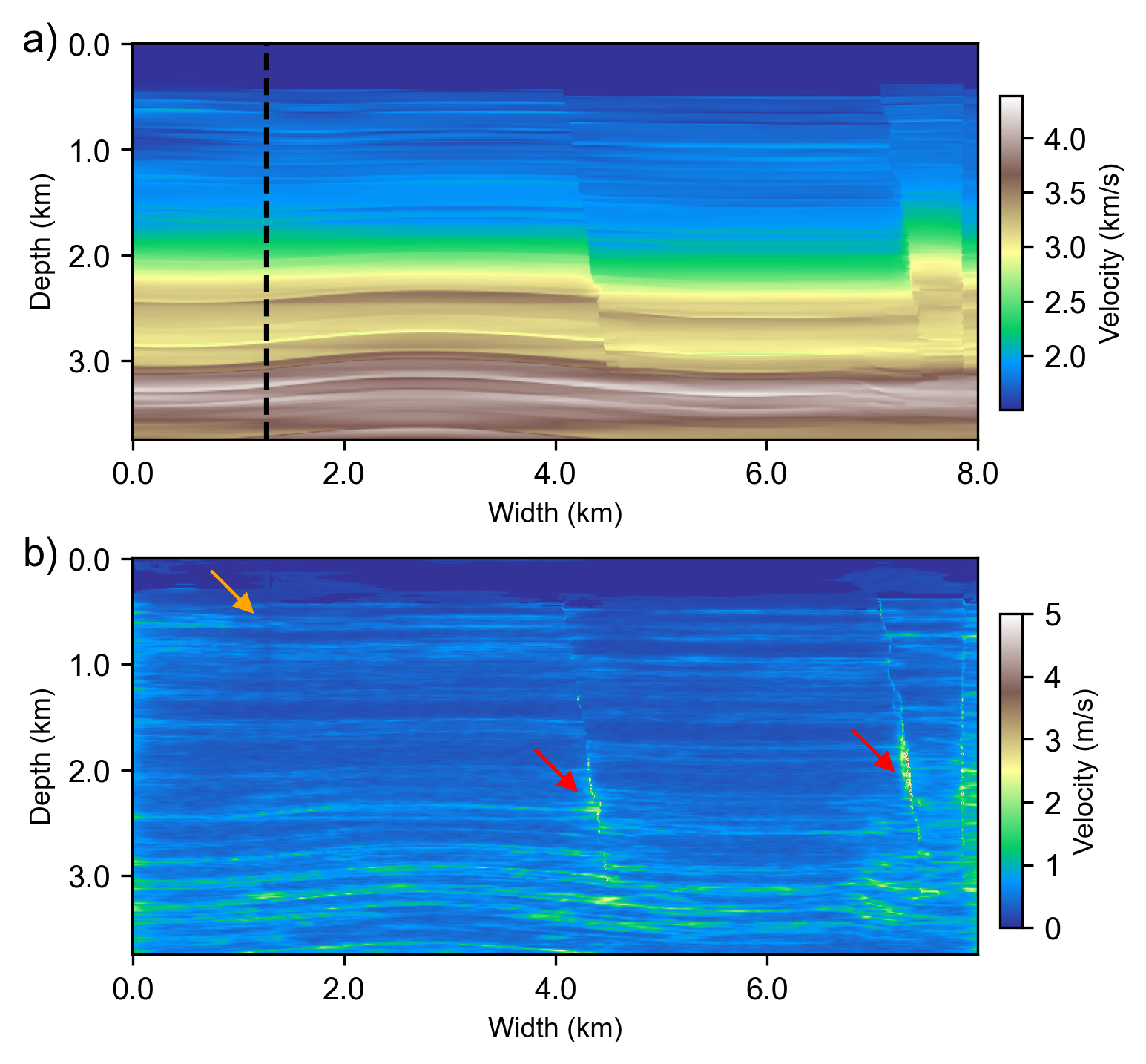}
\caption{Illustration of the uncertainty quantification using our method.(a) Mean velocity model computed from 50 realizations, with the well location marked by a dashed line.
(b) Standard deviation map of the 50 realizations.}
\label{fig16}
\end{figure}

To illustrate this capability, we generate 50 such samples for our target configuration and compute the mean and standard deviation across all realizations. 
Figs.~\ref{fig16}a and \ref{fig16}b show the resulting mean velocity and standard deviation, respectively. 
In panel (a), a dashed line indicates the well position. 
From the standard-deviation map in panel (b), we can observe three key features:
\begin{enumerate}
    \item \textbf{Reduced uncertainty near well control:} 
    At the well location (denoted by the orange arrow), the standard deviation is significantly lower, indicating that direct constraints from well data stabilize the velocity prediction.
    \item \textbf{Higher uncertainty around complex structures:} The model exhibits increased uncertainty near fault zones (red arrow), likely due to inaccuracies in the initial velocity model and the corresponding migrated images. As these constraints become less reliable, the generative process produces a wider range of velocity values.
    \item \textbf{Elevated uncertainty in deeper, high-velocity regions:} In areas of larger velocities at depth, the spread among the sampled models also grows.
\end{enumerate}

Overall, being able to perform uncertainty analysis during VMB offers valuable insights into the reliability of different subsurface regions. By sampling multiple models under the same well and imaging constraints, we can pinpoint where the model is well-constrained (e.g., near wells) and where additional data or improved imaging might reduce uncertainty (e.g., around faults or deeper high-velocity zones). This information can guide decision-making in subsequent exploration processes, such as planning new well locations or refining acquisition parameters.

\subsection{Limitations and Future work}
Despite the promising results, our work faces several notable limitations. First, although the generative model can learn diverse subsurface patterns, it is ultimately restricted by the synthetic training set it was exposed to. If the synthetic data distribution does not fully capture the complexities of a real geological environment, the generated velocity models may not generalize well to field data. Second, we have primarily validated our framework using well-controlled synthetic experiments, meaning its performance in actual field settings—where seismic data may have poor coverage or uncertain source wavelets—remains to be rigorously evaluated. Lastly, the diffusion model’s resolution constraints can become pronounced when scaling to larger or higher-resolution domains, where extrapolation beyond the trained grid size can reduce quality. For instance, a model trained on quarter-scale data might yield suboptimal reconstructions if directly applied to a domain four times larger.

Looking ahead, several strategies can help address these limitations. Future work could incorporate real well logs or partial field-data velocity models into the training phase, gradually bridging the gap between synthetic and actual geological conditions. Domain-adaptation techniques may also prove valuable, enabling a pre-trained synthetic model to be fine-tuned with a limited amount of field data. Moreover, certain validation against real seismic datasets would be essential for quantifying robustness and ensuring practical applicability. In terms of resolution, multi-scale training strategies or patch-based inference methods could be introduced, allowing the diffusion model to handle larger domains and finer details while preserving computational efficiency. 
\section{Conclusion}
We presented a generative diffusion approach for seismic velocity model building, designed to learn a prior distribution of subsurface velocities from key constraints, including well data, migrated images, and an initial model. A central aspect of our strategy lies in constructing a high-quality synthetic dataset that reduces generalization errors. By learning a prior distribution over the constructed realistic dataset, the framework provides a robust way to incorporate geological and structural constraints into the model-construction process. We benchmarked our method against conventional PWD-based well-interpolation techniques, which rely on smooth interpolation of well information across the model domain. While these approaches can provide reasonable large-scale trends, they often fail to capture sharp features like faults or rapid velocity transitions, especially in regions with sparse well control. In contrast, our method preserves such complex structures including the large-scale background information, so that mitigates local minima in full-waveform inversion (FWI), leading to higher-resolution reconstructions. Moreover, the probabilistic nature of the diffusion process allows for straightforward uncertainty quantification by sampling multiple velocity models, enabling practitioners to identify areas of low confidence.
\section*{Acknowledgments}
This work was supported in part by the National Natural Science Foundation of China (Grant No. 42274174 and 42130808) and in part by the Young Elite Scientists Program of State Key Laboratory of Geodesy and Earth’s Dynamics Grant No. S22L640102.

\bibliographystyle{IEEEtran}
\normalem    % Ignore “\usepackage{ulem}”
\bibliography{references}{}

\end{document}